\documentclass[twocolumn]{aastex61}

\newcommand{\HI}{\hbox{H~$\scriptstyle\rm I\ $}} 
\newcommand{\HII}{\hbox{H~$\scriptstyle\rm II\ $}} 
\newcommand{\gtsima}{$\; \buildrel > \over \sim \;$} 
\newcommand{\ltsima}{$\; \buildrel < \over \sim \;$}  
\newcommand{\gsim}{\lower.5ex\hbox{\gtsima}} 
\newcommand{\lsim}{\lower.5ex\hbox{\ltsima}} 
\newcommand{\msun}{\,{\rm \Msun}} 
\newcommand{\Msun}{\rm M_\odot}
\newcommand{\avchi}{$\langle \chi_{HI} \rangle$}

\received{2016 September 19}
\revised{2016 December 13}
\accepted{2017 January 10}
\submitjournal{ApJ}

\shorttitle{21cm-LAE cross correlations}
\shortauthors{Hutter et al.}
 
\begin{document}

\title{Exploring 21cm - Lyman Alpha emitter synergies for SKA}

\correspondingauthor{Anne Hutter}
\email{ahutter@swin.edau.au}

\author{Anne Hutter}
\affiliation{Swinburne University of Technology, Hawthorn, VIC 3122, Australia}
\affiliation{Leibniz-Institut f\"ur Astrophysik, An der Sternwarte 16, 14482 Potsdam, Germany}

\author{Pratika Dayal}
\affiliation{Kapteyn Astronomical Institute, University of Groningen, PO Box 800, 9700 AV Groningen, The Netherlands}

\author{Volker M\"uller}
\affiliation{Leibniz-Institut f\"ur Astrophysik, An der Sternwarte 16, 14482 Potsdam, Germany}

\author{Cathryn M. Trott}
\affiliation{International Centre for Radio Astronomy Research, Curtin University, Bentley WA 6103, Australia}
\affiliation{ARC Centre of Excellence for All-Sky Astrophysics (CAASTRO)}

\begin{abstract}

We study the signatures of reionization and ionizing properties of the early galaxies in the cross-correlations between the 21cm emission from the spin-flip transition of neutral hydrogen (\HI) and the underlying galaxy population, in particular a sub-population of galaxies visible as Lyman Alpha Emitters (LAEs). With both observables simultaneously derived from a $z\simeq6.6$ hydrodynamical simulation (GADGET-2) snapshot post-processed with a radiative transfer code (pCRASH) and a dust model, we perform a parameter study and aim to constrain both the average intergalactic medium (IGM)  ionization state ($1-\langle \chi_{HI} \rangle$) and the reionization topology (outside-in versus inside-out). We find that in our model LAEs occupy the densest and most-ionized regions resulting in a very strong anti-correlation between the LAEs and the 21cm emission. A 1000~h SKA-LOW1 - Subaru Hyper Suprime Cam experiment can provide constraints on $\langle \chi_{HI} \rangle$, allowing us to distinguish between IGM ionization levels of 50\%, 25\%, 10\% and fully ionized at scales $r\lsim 10$ comoving Mpc (assuming foreground avoidance for SKA). Our results support the inside-out reionization scenario where the densest knots (under-dense voids) are ionized first (last) for $\langle \chi_{HI} \rangle \gsim 0.1$. Further, 1000~h SKA-LOW1 observations should be able to confirm the inside-out scenario by detecting a lower 21cm brightness temperature (by about 2-10 mK) in the densest regions ($\gsim 2$ arcminute scales) hosting LAEs compared to lower-density regions devoid of them. 

\end{abstract}


% ***************************************************************************************************************
\section{Introduction}
% ***************************************************************************************************************

The Epoch of Reionization (EoR) marks the second, and last, major change in the ionization state of the Universe. The first stars and galaxies emit hydrogen ionizing photons that, permeate and, gradually ionize the vast majority of the the initially neutral hydrogen (\HI) in the intergalactic medium (IGM), marking the end of the EoR by $z\simeq6$ \citep{fan2006, ouchi2010, mcGreer2011, mcgreer2015}. However, the growth of ionized regions (the reionization topology) in the cosmic web and their dependence on the IGM over-density (whether reionization proceeded from over-to under-dense regions or vice versa) remain open questions to date. This is because both the progress and topology of reionization depend on a number of poorly understood parameters such as the abundance and spectral shapes of early galaxies, the fraction of ionizing photons produced by such sources that are able to escape galactic environment and contribute to reionization, and the IGM gas density distribution, to name a few.

Given that the key sources of ionizing radiation are located in high-density peaks, ionization fronts might be expected to originate in over-dense regions before percolating into under-densities. This scenario, referred to as ``inside-out" reionization, is supported by the majority of numerical and semi-numerical simulations \citep[e.g.][]{iliev2006a, iliev2012, trac2007, battaglia2013, bauer2015}. However, recombinations can outweigh ionization events in the densest regions of the IGM such that the outer parts of halos and filaments can become self-shielded and remain at least partially neutral. The existence of such self-shielded Lyman limit systems (LLS; sinks of Lyman-$\alpha$ photons), indicates that reionization may have an additional outside-in component \citep{miralda-escude2000,bolton2007,choudhury2009,bolton2013,kakiichi2016}, in which case ionizing photons might escape through low density tunnels into under-dense voids and ionize the over-dense filaments last \citep{finlator2009}. However, as these authors caution, a late reionization of filaments could also arise as a result of a highly biased emissivity field. 

Furthermore the nature of the key reionization sources, i.e. galaxies versus Active Galactic Nuclei (AGN), has recently come under discussion again. Over the past decade, a picture had emerged wherein star-forming galaxies were considered to be the main drivers of reionization \citep[e.g.][]{shapiro1987, choudhury2007, sokasian2003, sokasian2004, wyithe2003} with AGN contributing a negligible fraction to the total reionization photon budget \citep{fan2001,dijkstra2004, meiksin2005, bolton2007, srbinovsky2007, salvaterra2005, salvaterra2007, mcquinn2012}. 
However, with deep ($ -22.5\leq M_{1450} \leq -18.5$) observations of $z \simeq 4-6.5$ AGN, \citet{giallongo2015} find the faint end of the luminosity function to extend to two to four magnitudes fainter luminosities than that derived from previous surveys. The persistence of such high number densities of faint AGN to higher redshifts could imply AGN to be the main reionization drivers, with little/no contribution from galaxies \citep{madau2015}. Indeed, these authors show that an AGN-driven reionization scenario is quite capable of producing photoionization rates and an electron scattering optical depth in agreement with observations (of the Ly$\alpha$ forest and CMB polarization) and yields a reasonable reionization redshift of $z \simeq 5.7$.
Yet observations of \citet{giallongo2015} remain disputed as another analysis of the same field yields no convincing AGN candidates \citet{weigel2015}. Additionally, an AGN-driven reionization scenario is disfavored by the measured metal absorber abundances \citep{finlator2016} and Lyman-alpha forest measurements of the IGM temperature \citep{daloisio2016}.

Finally, the escape fraction of ionizing photons ($f_{esc}$) from galactic environments (the inter-stellar medium; ISM) into the IGM remains a debated quantity with theoretical estimates ranging from a few percent up to unity \citep[e.g.][]{ferrara2013, kimm2014, mitra2013}. Depending on the exact model used, its value either shows an increase \citep{gnedin2008, wise2009} or decrease \citep{razoumov2010, ferrara2013, wise2014, paardekooper2015} with halo mass or solely depends on redshift \citep[][and references within]{khaire2016}. In addition, (infrared) observations provide only weak constraints that are limited to galaxies at $z\simeq3-4$ \citep{cooke2014, vanzella2015}. 

Over the past few years high-$z$ Lyman Alpha Emitters (LAEs), detected through their Ly$\alpha$ emission (at 1216\AA\, in the rest-frame of the galaxy) have become popular probes of reionization. Given the sensitivity of Ly$\alpha$ photons to even trace amounts ($\simeq 10^{-5}$) of IGM \HI, a drop in the Ly$\alpha$ luminosity function (Ly$\alpha$ LF) accompanied by an increased clustering at $z \gsim 6.5$ has been interpreted to indicate an increase in the \HI fraction \citep{kashikawa2006, ouchi2010, kashikawa2011,mcquinn2007,jensen2013,mesinger2015,choudhury2015,castellano2016}. However, it must be noted that there are two additional effects that determine the ``observed" Ly$\alpha$ luminosity: firstly, the intrinsic luminosity produced depends on the fraction of \HI ionizing photons absorbed in the ISM ($1-f_{esc}$) that give rise to recombination lines including the Ly$\alpha$. Secondly, in addition to IGM attenuation, a fraction of the Ly$\alpha$ photons produced are absorbed by dust in the ISM \citep{dayal2008,dayal2011a,forero2010}. \citet{hutter2014,hutter2015} have shown that the effects of reionization, the ionizing photon escape fraction and dust are degenerate on the Ly$\alpha$ LF. Indeed, clustering information is required in order to be able to put additional constraints on the neutral fraction $\chi_{HI}$, since the decrease in the amplitude of the angular correlation function (ACF) is hard to attribute to anything other than reionization \citep[e.g.][]{mcquinn2007,jensen2013,hutter2015}. However it must be noted that, even combining all the available data sets (LFs+ACFs), LAEs can only shed light on the ``global" average IGM ionization state at any redshift. 

Radio interferometers including the forthcoming Hydrogen Epoch of Reionization Array \citep[HERA,][]{dillon2016} and the future Square Kilometer Array (SKA) aim to directly map out the reionization tomography by detecting the  21cm emission from the spin-flip transition of \HI. However, confirming the high-$z$ nature of the 21cm emission and interpreting the nature of reionization (inside-out versus outside-in) will require cross-correlating 21cm data with an unrelated data set \citep[e.g. high-z galaxies,][]{furlanetto-lidz2007}. The precise redshifts afforded by LAEs, in conjunction with the increasing number from new observations, renders them particularly attractive as one such data set. 
The 21cm-LAE cross-correlation has already been explored by a few works: using a combination of N-body and radiative transfer simulations, \citet{vrbanec2015} claim LOFAR should be able to detect an anti-correlation in the 21cm-LAE cross-correlation power spectrum on scales larger than $60 h^{-1}$Mpc; the signal is however dominated by LOFAR's system noise at smaller scales. Furthermore, using the 21CMFAST code \citep{mesinger2011} to model reionization on cosmological scales, \citet{sobacchi2016} show 1000~h observations with LOFAR should be able to distinguish at more than $1\sigma$ a fully ionized IGM from one that is half ionized at scales of about 3-10~Mpc. These authors find that the SKA phase 1 array will even be capable of distinguishing a fully ionized IGM from one than is a quarter ionized. However, both these models assume the emergent Ly$\alpha$ luminosity to effectively scale with the host halo mass. 

In this work, we pursue another approach by post-processing a $z\sim6.6$ hydrodynamic simulation snapshot (GADGET-2) that yield realistic galaxy populations with a dust model and a 3D radiative transfer code (pCRASH) to simultaneously derive the reionization topology and the underlying LAE distribution at $z \simeq 6.6$. Then, cross-correlating the 21cm signal with the LAE population for physical scenarios ($f_{esc}$, IGM ionization states and dust) in accord with LAE data (a) we show constraints that can be obtained on the IGM ionization state combining LAE and 21cm data, from the next generation Subaru and SKA observations, respectively, and (b) we show how these data sets can be used to answer the question of whether reionization had an inside-out, or outside-in topology.

We point out that this paper represents a parameter study to explore the signatures of properties of early galaxies on the ionization field, and in particular on the 21cm-galaxy cross correlation. In our model, we take the $z\simeq6.6$ output of a hydrodynamical simulation as a starting point of the evolution of the reionization field for 5 values of the escape parameter. Then we are able to (a) disentangle the effects of galactic properties on the 21cm-galaxy cross correlation from those entirely due to galaxy evolution, and (b) to evaluate the agreement of each physical scenario to LAE and 21cm observations at $z\sim6.6$. The next step is an post-processing of all our evolution steps of our hydrodynamical simulation for one realistic parameter set with transferring the ionization fields over the whole redshift range of reionization. This is planned for a future paper.

We start by describing our model for high-$z$ LAEs in Sec. \ref{sec_model}. We demarcate the location of the entire underlying galaxy population, and the fraction visible as LAEs as a function of the IGM density and ionization state in Sec. \ref{sec_gal_LAE_pdf}. We describe the characteristics of the 21cm-galaxy and 21cm-LAE cross-correlation in Sec. \ref{sec_link_21cm_gal}. We investigate the effects of reionization topology on the 21cm brightness temperature in over-densities/voids and provide estimates of the SKA-LOW1 detectability of the brightness temperature in regions with/without galaxies/LAEs in Sec. \ref{sec_reion_topology}, before concluding in Sec. \ref{sec_conclusions}.

% ***************************************************************************************************
\section{The model}
\label{sec_model}
% ***************************************************************************************************

In this Section we describe our model for $z \simeq 6.6$ LAEs that combines a cosmological smoothed particle hydrodynamic (SPH) simulation run using GADGET-2 \citep{springel2005} with the pCRASH radiative transfer (RT) code \citep{partl2011}  and a model for ISM dust \citep{dayal2010}. The interested reader is referred to \citet{hutter2014, hutter2015} for a detailed description.

The hydrodynamical simulation used as the basis for our model is run with the TreePM-SPH code GADGET-2 with a box size of $80h^{-1}$~comoving Mpc (cMpc). The simulation follows a total of $2\times1024^3$ dark matter (DM) and gas particles, resulting in a DM and gas particle mass resolution of $3.6\times10^7h^{-1}\Msun$ and $6.3\times10^6h^{-1}\Msun$, respectively. It includes physical prescriptions for star formation, metal production and feedback as described in \citet{springel2003}, assuming a \cite{salpeter1955} initial mass function (IMF) between $0.1-100\Msun$. Bound structures with more than 20 particles are identified as galaxies using the Amiga Halo Finder \citep[AHF;][]{knollmann2009}. We only use ``resolved" galaxies leading to a complete halo mass function, with at least 10 star particles (corresponding to a minimum of 160 gas particles) and a halo mass $M_h>10^{9.2}\Msun$. We obtain the total intrinsic spectrum for each galaxy summing over all its star particles using the stellar population synthesis code STARBURST99 \citep{leitherer1999}; the intrinsic spectrum for each star particle naturally depends on its mass, age and metallicity. For each galaxy we compute the dust mass produced considering Type II SN (SNII) to be the main dust factories in the first billion years; the corresponding UV attenuation is calculated using the dust model described in \citet{dayal2010}. 

The observed ultra-violet (UV) luminosity ($L_c^{obs}$) is then related to the intrinsic value ($L_c^{int}$) as $L_c^{obs}=f_c\times L_c^{int}$, where $f_c$ is the fraction of UV continuum ($\sim1500$\AA) photons that escape the ISM unattenuated by dust. Further, the observed Ly$\alpha$ luminosity is calculated as $L_{\alpha}^{obs}=L_{\alpha}^{int} f_{\alpha} T_{\alpha}$ where $f_\alpha$ and $T_{\alpha}$ account for the Ly$\alpha$ attenuation by ISM dust and IGM \HI, respectively. In accord with observational selection criteria, galaxies with an absolute UV magnitude $M_{UV}\lsim-17$ are identified as Lyman break Galaxies (LBGs); galaxies with $L_\alpha^{obs} \geq 10^{42}$~erg~s$^{-1}$ and a Ly$\alpha$ equivalent width $EW = L_\alpha^{obs}/L_c^{obs} \geq 20$\,\AA\, are identified as LAEs. We note that LAEs are a subset of the underlying LBG population: whilst only a fraction of LBGs show Ly$\alpha$ emission fulfilling the observational criterion, depending on the IGM ionization state and dust clumping in the ISM, all LAEs are bright enough in the UV to be classified as LBGs \citep[see e.g.][]{dayal2012, hutter2015}.

In order to obtain $T_\alpha$, we post-process the $z\simeq6.6$ snapshot of our hydrodynamical simulation with the RT code pCRASH \citep{partl2011}. pCRASH is a MPI\footnote{http://www.mpi-forum.org} parallelized  version of CRASH \citep{ciardi2001,maselli2003,maselli2009} which is a 3D RT code capable of treating multiple source spectra, a spatially dependent clumping factor and evolving density fields. 
pCRASH naturally yields both the evolving ionization fields and the IGM temperature that are used to calculate the 21cm \HI emission, as explained in Sec. \ref{subsec_21cm_over-dense_regions}. 
Given the poor constraints on the escape fraction of ionizing photons $f_{esc}$ ($\lambda<912$\AA), and its impact on both the IGM ionization state as well as the intrinsic Ly$\alpha$ luminosity, we explore a wide range of values such that $f_{esc}=0.05$, $0.25$, $0.5$, $0.75$, $0.95$.
Thus, we perform $5$ RT simulations, whereas in each a different global $f_{esc}$ value is assumed for all galaxies.
Starting from a completely neutral IGM, we run pCRASH until the IGM is fully ionized, reaching an average \HI fraction of \avchi $\simeq 10^{-4}$, for each $f_{esc}$ value by following the ionizing radiation from $31855$ ``resolved" galaxies on a $128^3$ grid.
With pCRASH computing the evolution of the ionized regions, we obtain the ionization history, i.e. snapshots at different \avchi, for each of our chosen $f_{esc}$ values.
Assuming a Gaussian profile for the Ly$\alpha$ line emerging from the galaxy, we derive $T_{\alpha}= e^{-\tau_{\alpha}}$ by averaging the IGM attenuation along $48$ random lines of sight (LOS) for each galaxy, with $\tau_{\alpha}$ being the optical depth to \HI along the LOS. 
Once $T_{\alpha}$ is calculated for each combination of $f_{esc}$ and \avchi, the only free parameter left to fit model results to LAE observables (Ly$\alpha$ LF, LAE angular correlation functions) is $f_\alpha$ - we parameterize this as the ratio between the escape fractions of Ly$\alpha$ and UV continuum photons, $p=f_{\alpha}/f_c$. 

Matching the theoretical LAE Ly$\alpha$ LF to observations \citep{kashikawa2011}, we uncover a three-dimensional degeneracy between $f_{esc}$, \avchi~and $f_\alpha/f_c$ such that the data is equally well fit for \avchi$\simeq0.5-10^{-4}$, $f_{esc}\simeq0.05-0.5$  and $f_{\alpha}/f_c=0.6-1.8$ within a $1\sigma$ error. Physically this implies that a decrease in $T_\alpha$ (due to a more neutral IGM) can be compensated by a larger escape through the ISM. Folding in the ACF data \citep{kashikawa2011}, we find that LAE clustering is extremely sensitive to the reionization state with large-scale clustering signatures being impossible to attribute to anything other than the IGM ionization state. Indeed, adding ACF constraints and allowing for clumped dust ($f_\alpha/f_c\gsim0.6$) yields constraints of \avchi$\lesssim0.25$ within a $4\sigma$ error at $z\simeq6.6$.

We note that a different - non Gaussian, e.g. double wing - line profile yields consistent results, as a potential overall increase in $T_{\alpha}$ is compensated by a lower $f_{\alpha}$ \citep[for a detailed discussion see][or Appendix \ref{a2_sec_lya_line_profile}]{hutter2014}.

% ***************************************************************************************************
\section{Distribution of galaxies and 21cm in over-dense and ionized regions}
\label{sec_gal_LAE_pdf}
% ***************************************************************************************************

Now that the galaxy populations visible as LAEs and LBGs have been identified we study the probability density distribution of all galaxies and the subset visible as LAEs as a function of over-density and the surrounding IGM ionization state in Sec. \ref{subsec_gal_tracer}. We then discuss how the presence of galaxies in over-densities impacts the probability density distribution of the 21cm brightness temperature in Sec. \ref{subsec_21cm_over-dense_regions}.

% ***************************************************************************************************
\subsection{Galaxies as tracers of ionized and over-dense regions}
\label{subsec_gal_tracer}
\begin{figure}
 \centerline{\includegraphics[width=0.55\textwidth]{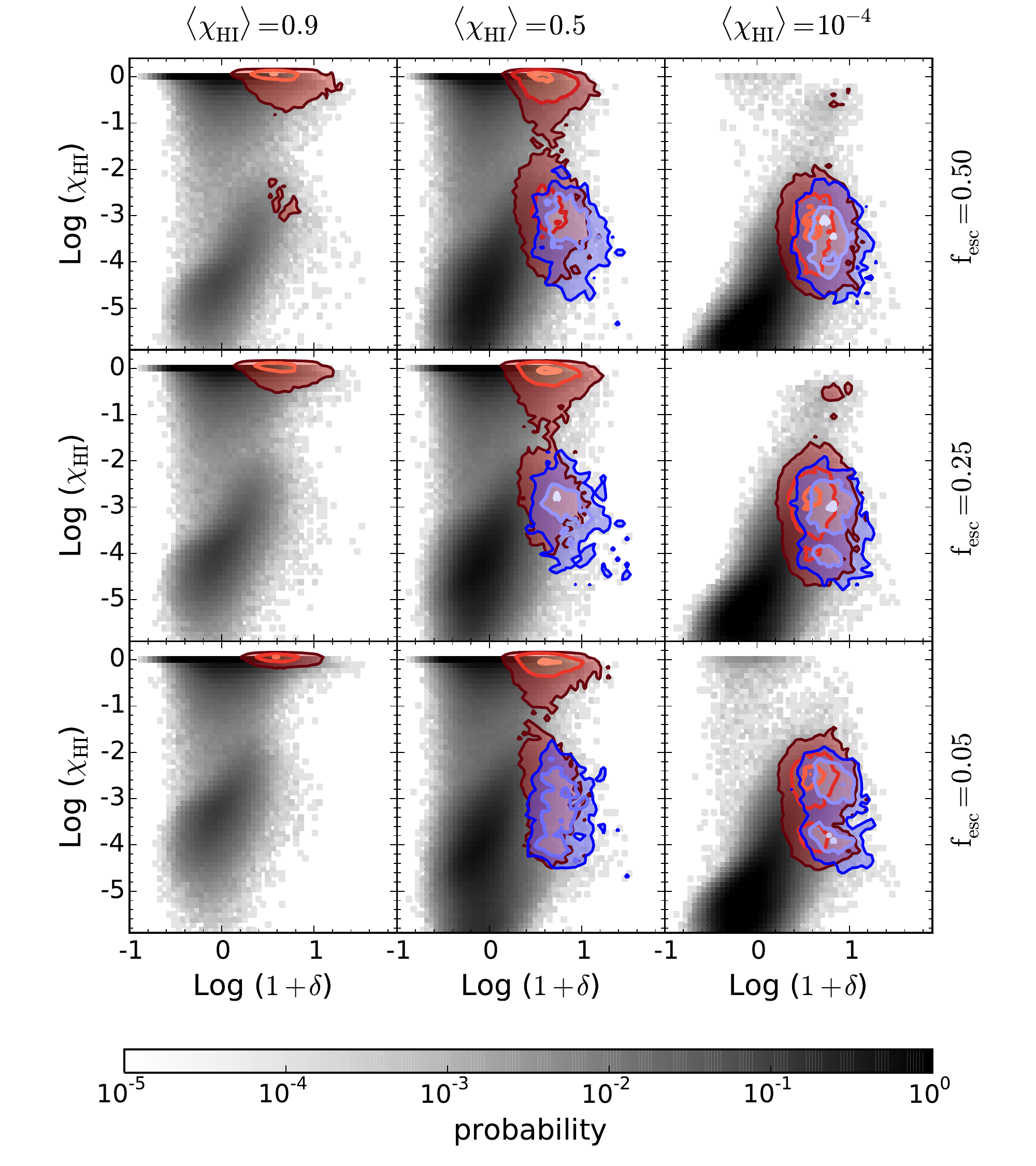}}
 \caption{Probability density distribution of the IGM gas (drawn from all simulation cells) as a function of gas over-density and neutral hydrogen fraction (gray color scale) for three different $f_{esc}$ ($=0.05$, $0.25$, $0.5$; rows) and \avchi\, ($=0.9$, $0.5$, $10^{-4}$; columns) combinations. The dark red, red and light red contours show the regions occupied by 10\%, 50\% and 90\% of all galaxies, respectively. Dark blue, blue and light blue contours show the regions occupied by 10\%, 50\% and 90\% of LAEs, respectively; no galaxies are visible as LAEs for \avchi~$\simeq 0.9$. \label{fig_dens_XHI}}
\end{figure}

% ***************************************************************************************************
We use the $f_{esc}$ and $\langle \chi_{HI} \rangle$ combinations identified (in Sec. \ref{sec_model}) to analyze the relation between the IGM gas over-density/ionization state and the underlying galaxy population. 
We start by discussing the probability density distribution of the neutral hydrogen fraction ($\chi_{HI}$) with respect to the gas over-density ($1+\delta$) in the entire simulation box and compare it to cells containing galaxies before studying how these differ in the subset of galaxies visible as LAEs; we explore all parameter ($f_{esc}$, $f_\alpha/f_c$ and \avchi) combinations that are in accord with the observed LAE LFs \citep[see][]{hutter2014}. 
We use each computing cell of the grid ($1+\delta$, $\chi_{HI}$) in our RT calculations to derive the probability density distributions of the IGM gas in the entire simulation box, which are shown by means of the gray scale in Fig. \ref{fig_dens_XHI} for $f_{esc}=0.05$, $0.25$, $0.5$ (rows) and \avchi$\simeq0.9$, $0.5$, $10^{-4}$ (columns).
With a size of $625 h^{-1}$ckpc, our simulated galaxies mostly lie within one computing cell. Due to this rather coarse resolution we omitted subtracting galaxies from the gas density grid. Only a small fraction of cells contain galaxies and their circumgalactic medium, while the majority of cells represent the IGM. For convenience we refer always to the IGM in the following.

We expect the bulk of the cells to be completely neutral in the initial stages of reionization. The successive growth and overlap of ionized regions would lead to an increase in the local photoionization rate, resulting in shifting the bulk of cells towards lower $\chi_{HI}$~ values. Finally, we expect very few cells to have a neutral fraction $\chi_{HI}>10^{-3}$ for a fully ionized universe. This is exactly the behavior shown by our results in Fig. \ref{fig_dens_XHI}. 

As expected, we find galaxies to lie in regions of over-density as seen from the red contours in Fig. \ref{fig_dens_XHI}: while the least massive galaxies lie in marginally over-dense regions ($1+\delta \simeq 1.5$), the most massive galaxies lie in regions 10 to 15 times more over-dense than average. While the over-density is fixed by the SPH simulation, the distribution of galaxies in the ionization field naturally evolves as reionization progresses. 
 From the sharp ionization fronts of stellar sources, we expect a bimodal probability distribution in $\chi_{HI}$ distinguishing the ionized ($\chi_{HI}\lesssim0.01$) from the neutral ($\chi_{HI}=1$) regions. Indeed we see from Fig. \ref{fig_dens_XHI} that most cells are either ionized or neutral. However, we also find partially ionized cells ($0.01\lesssim\chi_{HI}<1$), whose existence is a consequence of the finite resolution of our RT simulations.
In this context we can understand galaxies being located in neutral/partially ionized cells as sources that have not emitted enough photons to fully ionize their cell. Thus, given their large over-densities, most galaxies lie in neutral cells in the initial stages of reionization (\avchi$\simeq 0.9$). 
As expected, the distribution of galaxies widens considerably for a half ionized IGM: while some galaxies still occupy neutral regions ($\chi_{HI} \simeq 1$), others (possibly those in clustered regions) are embedded in a fully ionized IGM with $\chi_{HI} \simeq 10^{-4}$. Finally, the distribution of all galaxies shifts to lie at $\chi_{HI} \lsim 10^{-2}$ for a fully ionized IGM. At a given IGM state (columns in the same figure), differences in galaxy distributions naturally vary with $f_{esc}$ since the strength of the photoionization field created by any source scales with this parameter. It is interesting to note that, due to a combination of low IGM densities and photoionization rate contributions from multiple galaxies, many under-dense regions are as highly ionized as $\chi_{HI} \simeq 10^{-5}$ even for an average ionization state of \avchi$\simeq 0.9$. 

We find that the observed Ly$\alpha$ LF at $z\sim6.6$ can only be reproduced for \avchi$\lsim 0.5$; for \avchi$>0.5$ the number of galaxies identified as LAEs drops significantly, since the IGM Ly$\alpha$ transmission decreases considerably for rising \avchi~values.
We find that the subset of galaxies visible as LAEs (blue contours in Fig. \ref{fig_dens_XHI}) are those that lie in the most over-dense (over-density between $2$ and 15) and highly ionized regions ($\chi_{HI}\lsim 10^{-2}$). Although the galaxy population is not evolving in our scenario, we expect our findings to be robust given the conditions required for galaxies to be visible as LAEs: firstly, galaxies must produce enough intrinsic Ly$\alpha$ luminosity and secondly, they must transmit enough of this luminosity through the IGM so as to result in $L_\alpha^{obs} \gsim 10^{42}{\rm erg\, s^{-1}}$. Given that the spatial scale imposed by the Gunn-Peterson damping wing on the size of the \HII region corresponds to a redshift separation of $\Delta z \approx 4.4 \times 10^{-3}$, i.e. about 280 kpc (physical) at $z=6$ \citep{miralda-escude1998}, $z \simeq 6.6$ LAEs require a halo mass $\gsim 10^{9.5}\Msun$ \citep{dayal2012}, with significantly larger ($10^9-10^{11}\msun$) stellar masses being inferred observationally \citep{pentericci2009}. Naturally, as the high-mass end of the galaxy population, LAEs are expected to lie in the most over-dense and highly ionized regions.

% ***************************************************************************************************
\subsection{21cm emission from over-dense neutral regions}
\label{subsec_21cm_over-dense_regions}
% ***************************************************************************************************

\begin{figure}
 \centerline{\includegraphics[width=0.55\textwidth]{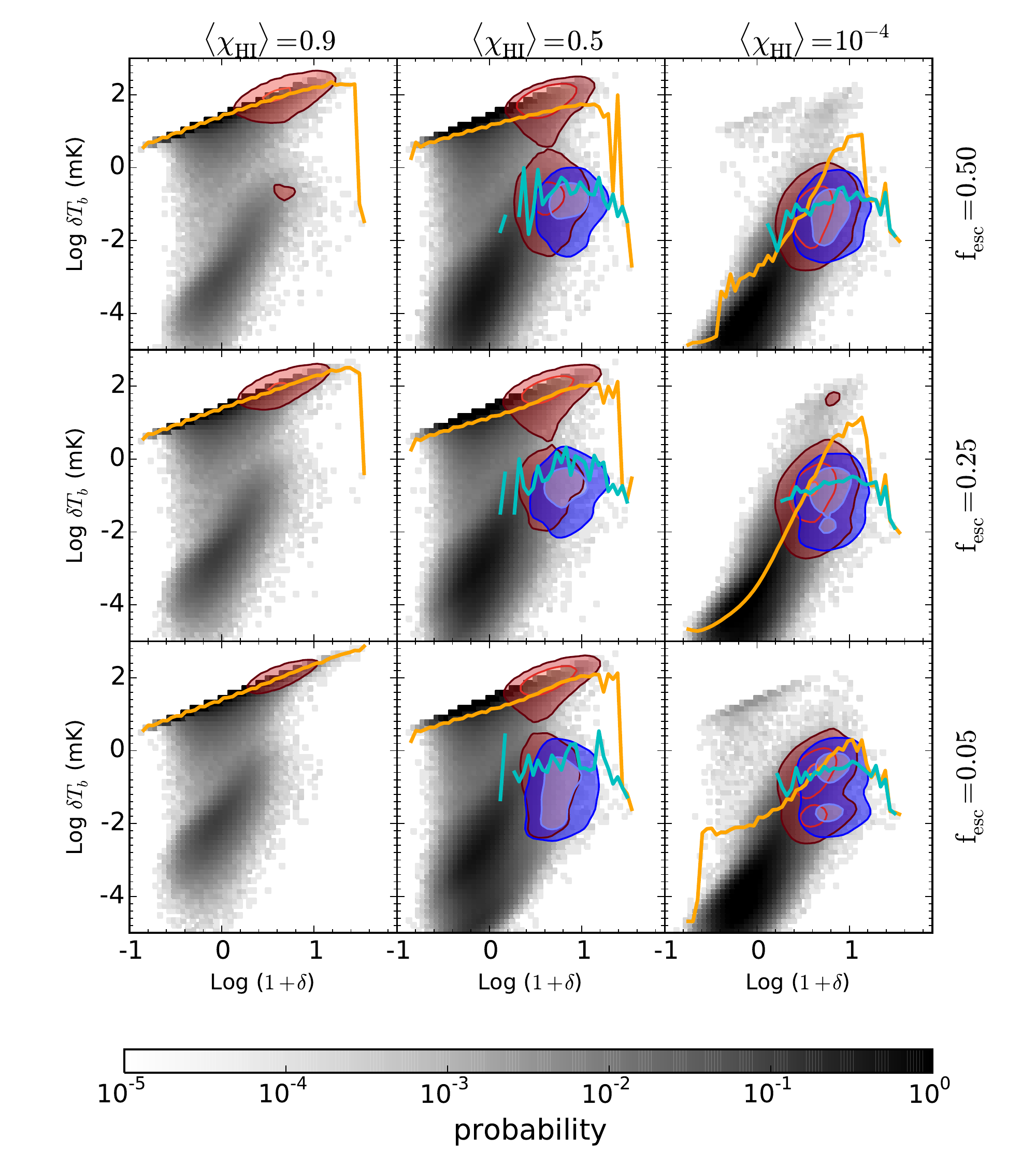}}
 \caption{Probability density distribution of the IGM gas as a function of gas over-density ($1+\delta$) and the 21cm differential brightness temperature ($\delta T_b$) for three different $f_{esc}$ ($=0.05$, $0.25$, $0.5$; rows) and \avchi\, ($=0.9$, $0.5$, $10^{-4}$; columns) combinations. The dark red, red and light red contours show the regions occupied by 10\%, 50\% and 90\% of all galaxies, respectively. Dark blue, blue and light blue contours show the regions occupied by 10\%, 50\% and 90\% of LAEs, respectively. The thick solid orange line shows the mean value of $\delta T_b$ for all cells; the blue line shows the much lower mean $\delta T_b$ value in cells hosting LAEs.  \label{fig_fracdens_Tb}}
\end{figure}

The distribution of the neutral hydrogen in the IGM can be observed through its 21cm brightness temperature, which measures the intensity of the emission (or absorption) of 21cm radiation against the Cosmic Microwave Background (CMB). We now discuss how the 21cm brightness temperature depends on the IGM density and its dependence on the presence of galaxies, specially the subset visible as LAEs. We start by calculating the differential 21cm brightness temperature ($\delta T_b$) in each of the ($128^3$) pCRASH cells as \citep[e.g.][]{iliev2012}
\begin{eqnarray}
 \delta T_b (\vec{x}) &=& T_0\ \langle \chi_{HI} \rangle \ \left(1+\delta(\vec{x})\right) \ \left(1+\delta_{HI}(\vec{x})\right),
 \label{eq_21cm_brightness_temperature}
\end{eqnarray}
where
\begin{eqnarray}
 T_0&=&28.5 \mathrm{mK}\ \left( \frac{1+z}{10} \right)^{1/2} \frac{\Omega_b}{0.042} \frac{h}{0.073} \left( \frac{\Omega_m}{0.24} \right)^{-1/2}.
 \label{eq_T0}
\end{eqnarray}
Here, $\Omega_b$ and $\Omega_m$ represent the baryonic and matter densities respectively and $h$ is the Hubble parameter. Further, $1+\delta(\vec{x})=\rho(\vec{x})/\langle\rho\rangle$ represents the local gas density compared to the average global value and $1+\delta_{HI}(\vec{x})=\chi_{HI}(\vec{x})/\langle \chi_{HI}\rangle$ represents the local \HI density fraction compared to the average global value. 
Our computation of the differential 21cm brightness temperature does not include fluctuations in the spin temperature and peculiar velocities of the gas. For \avchi$\lesssim0.8$, spin temperature fluctuations become negligible, as the heating of the IGM by X-rays from the first sources leads to spin temperatures that exceed well the CMB temperature \citep{ghara2015}. Spin temperature fluctuations may only become important when the IGM is mostly neutral and has not been entirely preheated. Similarly, the effect of peculiar velocities is only imprinted in the 21cm power spectrum as long as the 21cm signal is not dominated by the \HI fluctuations, e.g. at high \avchi~values \citep{ghara2015}.

In Fig. \ref{fig_fracdens_Tb} we show the resulting probability density distribution in terms of the over-density and the differential 21cm brightness temperature. As seen from Eqn. \ref{eq_21cm_brightness_temperature}, $\delta T_b$ depends both on the gas over-density as well as the neutral fraction that together determine the \HI density in any cell. The fact that most over-dense cells contain \HI with only a few cells being ionized in the initial reionization stages ($ \langle \chi_{HI} \rangle\simeq 0.9$; sec. \ref{subsec_gal_tracer} above), results in the average $\delta T_b$ (orange line) increasing from about 4 to 250~mK as the density increases from $1+\delta = 0.14$ to $14$ times the average density. As reionization progresses to a state where it is half completed ($ \langle \chi_{HI} \rangle \simeq 0.5$), the distribution of all cells shifts towards more ionized values, resulting in a drop in the average $\delta T_b$ amplitude, ranging between 4 to 200~mK from the rarest to the densest regions. Finally, given that very few cells have $\chi_{HI} \gsim 10^{-2.5}$ for a fully ionized IGM ($ \langle \chi_{HI} \rangle\simeq 10^{-4}$), the average $\delta T_b$ signal changes both in shape and amplitude although it still increases with the over-density given that the \HI density scales with this. Naturally, however, $\delta T_b$ has much lower values ranging between about $10^{-4.5}$ to 10~mK. For each of the $\langle \chi_{HI} \rangle$-$f_{esc}$ combinations studied here, the drop in $\delta T_b$ for $1+\delta \gsim 14$ values is essentially a statistical fluctuation driven by a few over-dense cells.

Given that galaxies are located in highly over-dense and neutral regions in the early stages of reionization, simulation cells hosting galaxies show a high $\delta T_b$ signal ($\simeq 4-150$~mK). As reionization proceeds to being half complete, the galaxy distribution widens considerably - while a fraction of galaxies occupy ionized regions with $\delta T_b \sim 0.003$ to 4~mK, field galaxies are embedded in partly neutral regions showing $\delta T_b \sim 4$ to 250~mK. Finally, the $\delta T_b$ signal from cells hosting galaxies drops by about two orders of magnitude to 0.003 to a few mK once reionization is complete. 

\begin{figure*}
 \center{\includegraphics[width=0.9\textwidth]{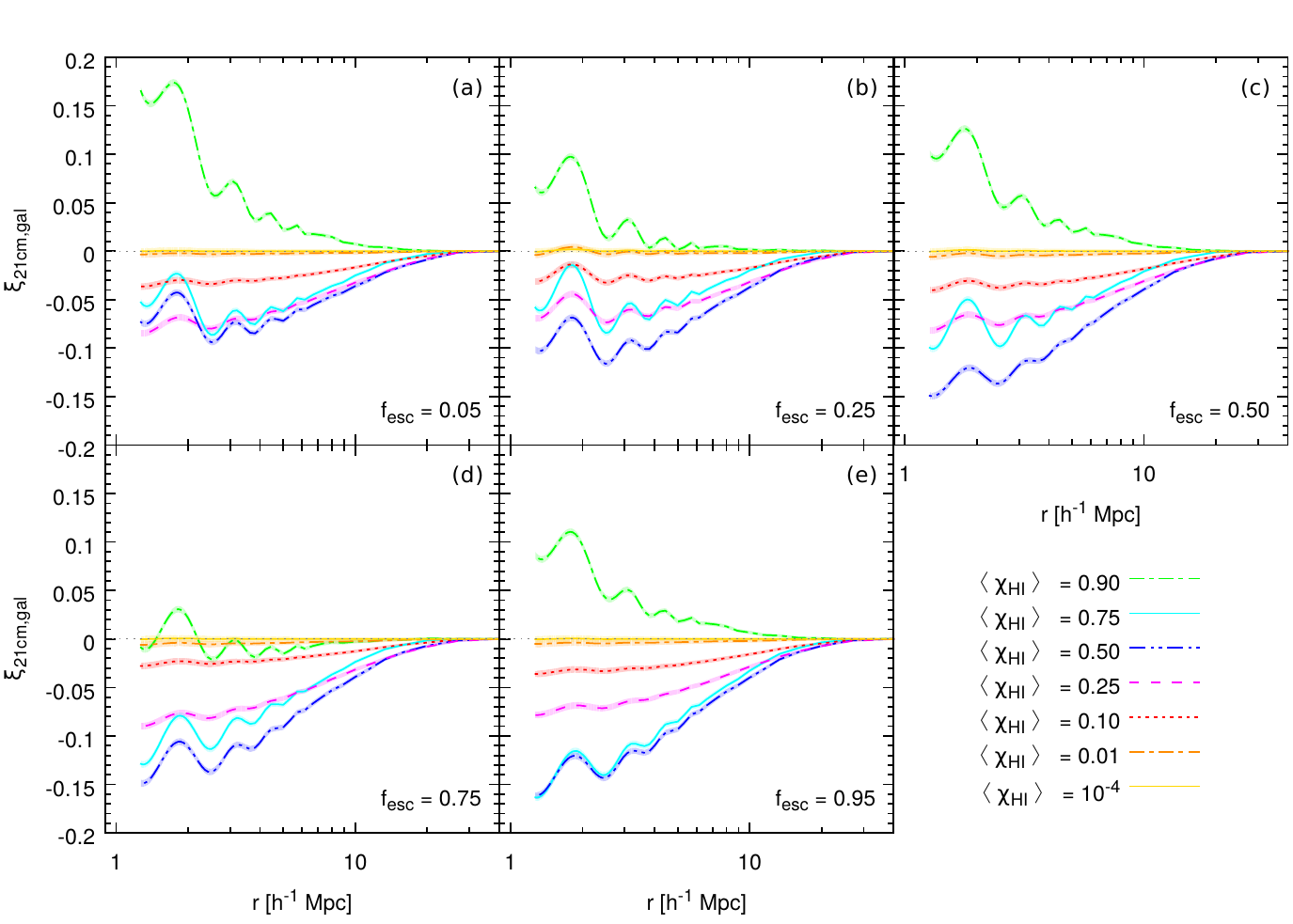}}
 \caption{The cross-correlation function $\xi_{21cm,gal}$ for the differential 21cm brightness temperature and galaxy distribution as a function of distance $r$ from galaxies. In each panel, lines indicate $\xi_{21cm,gal}$ for different IGM ionization states, ranging between \avchi$=0.9$ - $10^{-4}$; shaded regions show the uncertainties associated with SKA1-Low 1000~h observations. As marked, panels show results for different $f_{esc}$ values ranging between $0.05$ and $0.95$. See text in Sec. \ref{subsec_21cm_gal_crosscorrelation}.\label{fig_21cm_gal_crosscorrelation}}
\end{figure*}

Given that only highly clustered sources in ionized regions are visible in the Ly$\alpha$ in the initial reionization stages \citep{mcquinn2007, hutter2015}, LAEs exhibit much lower average $\delta T_b$ values (blue line in Fig. \ref{fig_fracdens_Tb}) compared to cells without galaxies at the same over-density: e.g. LAE hosting cells show $\delta T_b \simeq 1$ mK compared to $\simeq 10$ mK shown for $1 + \delta \simeq 3$ for a half-ionized IGM; we remind the reader no galaxies are visible as LAEs for $\langle \chi_{HI} \rangle \simeq 0.9$. Finally, the signal from LAEs is quite similar to that from galaxies, ranging between 0.003 to a few mK once reionization is complete. However, given that they preferentially occupy ionized regions compared to the entire underlying galaxy population, LAE hosting cells show much less variation in the average $\delta T_b$ compared to that averaged over all simulation/galaxy hosting cells.

% ***************************************************************************************************
\section{Linking 21cm emission to the underlying galaxy population}
\label{sec_link_21cm_gal}
% ***************************************************************************************************

Differential 21cm brightness temperature tomographic maps will be ideal indicators of the IGM ionization history. However understanding the key reionization sources, and indeed even verifying that the 21cm signal originates at high-redshifts, will require correlating the 21cm brightness temperature signal with the underlying galaxy population and specially the subset visible as LAEs given their precise redshifts \citep[e.g.][]{furlanetto-lidz2007, lidz2009, wyithe2007}. In this section, we show the IGM ionization state constraints possible with the SKA by combining 21cm-galaxy data (Sec. \ref{subsec_21cm_gal_crosscorrelation}) and 21cm-LAE data (Sec. \ref{subsec_21cm_LAE_crosscorrelation}).

% ***************************************************************************************************
\subsection{The 21cm - galaxy cross-correlation}
\label{subsec_21cm_gal_crosscorrelation}
% ***************************************************************************************************

Taking $\Delta_i$ and $\Delta_j$ to represent the 21cm brightness temperature and galaxy fields, respectively, their cross-correlation power spectrum can be calculated as \citep{park2014}
\begin{eqnarray} 
P_{i,j}(k) &=& \langle \tilde{\Delta}_i (\vec{k})\ \tilde{\Delta}_j(-\vec{k}) \rangle,\\ 
\tilde{\Delta_l}(\vec{k})&=&\frac{1}{V} \int \Delta_l(\vec{x})\ \exp(-i\vec{k}\vec{x})\ \mathrm{d}^3x\ \ \mathrm{for}\ l=i,j. 
\end{eqnarray} 
Computing the Fourier transformation of the cross power spectrum $P_{21cm,gal}(k)$, we derive the cross correlation function as 
\begin{eqnarray}
\xi_{21cm,gal}(r)=\frac{1}{(2\pi)^3} \int  P_{21cm,gal}(k)\ \frac{\sin(kr)}{kr}\ 4\pi k^2\ \mathrm{d}k.
\label{eq_crosscorrfunc}
\end{eqnarray}
The integration along the k-axis is carried out numerically using the composite trapezoidal rule. The resulting cross correlation functions are shown in Fig. \ref{fig_21cm_gal_crosscorrelation} for $f_{esc}$ values ranging between $0.05$ to $0.95$ and the associated ionization states ranging from fully neutral to fully ionized. Throughout this paper we compute the cross correlation functions taking the entire simulation box volume into account. Our simulated volume is sufficient to trace the cross correlations up to a scale of $\sim10h^{-1}$~cMpc. The finite box size poses a lower limit in k-space ($k<k_{\mathrm{lim}}\simeq 0.2 h$~cMpc$^{-1}$), which introduces uncertainties in the amplitude of our cross correlation function. However, given the cross power spectrum has the same sign at $k<k_{\mathrm{lim}}$ than at $k_{\mathrm{lim}}$, an extension of the integration in Equation \ref{eq_crosscorrfunc} to lower $k_{\mathrm{lim}}$ values would result in higher amplitudes in the cross correlation function (indicating an even stronger correlation/anti-correlation for a positive/negative $P_{21cm,gal}(k)$).

In the early stages of reionization (\avchi$\simeq0.9$, beginning of RT simulation), the 21cm brightness temperature and galaxy distribution are positively correlated on small ($\lsim 10 h^{-1}$ cMpc) scales as seen in Fig. \ref{fig_21cm_gal_crosscorrelation}. This behavior is driven by galaxies being embedded in over-dense and only partly ionized regions that therefore show 21cm emission, as also seen from Fig. \ref{fig_fracdens_Tb}. As expected the strength of the correlation decreases with increasing scale, saturating to 0 at $\simeq 20 h^{-1}$~cMpc where galaxy positions and 21cm emission are uncorrelated. As the global neutral hydrogen fraction drops to \avchi$\gsim 0.75$, the correlation flips in sign and becomes anti-correlated at small-scales - this is driven by galaxies being embedded in mostly ionized regions. The strength of the anti-correlation is strongest for \avchi$\simeq0.5$ and then decreases with decreasing \avchi\, as the \HI content becomes lower, leading to the 21cm emission approaching 0 - this happens for \avchi$\simeq 0.01$, irrespective of the parameters used. We find that $\xi_{21cm,gal}$ shows significant small-scale fluctuations for \avchi$\gsim0.5$ where (low-mass, $M_{\star}\lesssim 10^{9.5}\Msun$) galaxies are embedded in partly ionized regions of varying sizes. Indeed, we find that the relative amplitude of the oscillations decreases as the IGM becomes more ionized, resulting in field/low-mass galaxies being enclosed by increasingly ionized regions. Given that we use a single simulation snapshot, the underlying galaxy field is fixed in this work and, we note that the variations in $\xi_{21cm, gal}$ are solely introduced by an evolution of the ionization fields. 

We find that these above results qualitatively hold true for all the $f_{esc}$ values explored in this work. We remind the reader that the average ionization fraction is a combination of the volume ionized and the degree of ionization: an increasing $f_{esc}$ leads to a higher degree of ionization, for a given galaxy population, requiring smaller ionized volumes to result in a given \avchi~\citep[e.g.][]{hutter2014} - this results in the slight differences in $\xi_{21cm,gal}$ with varying $f_{esc}$ for a given \avchi. We therefore find the strongest small-scale anti-correlation for $f_{esc}=0.95$ where the \HI is most highly ionized; the correlation strength decreases with decreasing $f_{esc}$ where galaxies reside in larger ionized regions for a given \avchi.

We then calculate the ability of SKA1-Low to discriminate between models, by computing the $1\sigma$ uncertainties on the 21cm-galaxy cross correlation functions for an idealized 1000~h SKA1-Low experiment. The thermal noise and sample variance include the most recent array configuration V4A\footnote{http://astronomers.skatelescope.org/wp-content/uploads/2015/11/SKA1-Low-Configuration\_V4a.pdf}, with a filling factor that reduces substantially outside the core, yielding poorer brightness temperature sensitivity performance on small scales. The system temperature and effective collecting area as a function of frequency are matched to the systemic specifications in the {\it SKA1 System Baseline Design} document\footnote{http://astronomers.skatelescope.org/wp-content/uploads/2016/05/SKA-TEL-SKO-0000002\_03\_SKA1SystemBaselineDesignV2.pdf}.
We assume no foreground subtraction (foreground avoidance), where the foregrounds dominate the 21cm signal in an extended k-space region (''wedge''). We pursue a rather pessimistic approach, where the available k-space is reduced by a factor of two.
The bandwidth is matched to each distance, $r$, in the two-point cross-correlation function, with a minimum resolution of 1.9$h^{-1}$~Mpc.
As shown in Fig. \ref{fig_21cm_gal_crosscorrelation}, we again encounter a degeneracy between $f_{esc}$ and \avchi: we are unable to discriminate between IGM ionization states of \avchi$\gsim 0.1$ for $f_{esc}=0.05-0.95$. The only ionization state that could be unambiguously observed with the SKA1 corresponds to the positive, albeit fluctuating, correlation seen for an almost neutral IGM with \avchi$> 0.9$.

\begin{figure*}
 \center{\includegraphics[width=0.9\textwidth]{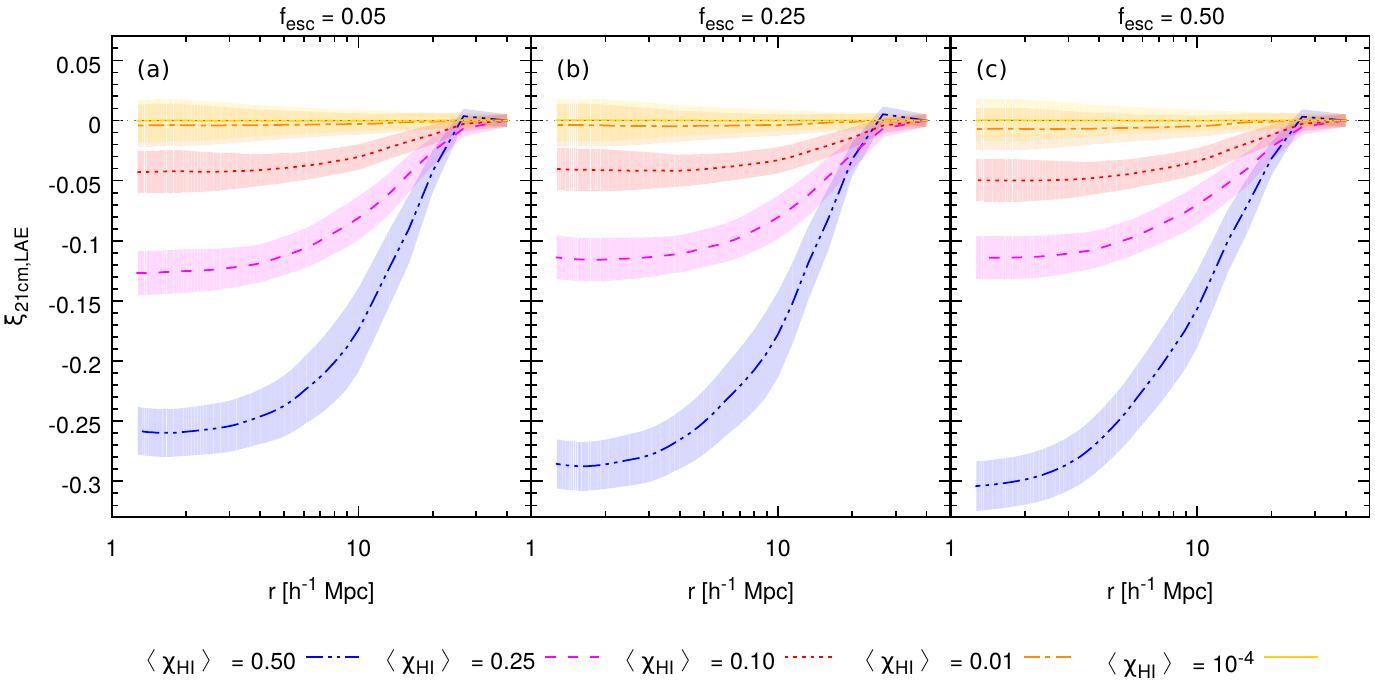}}
 \caption{The cross-correlation function $\xi_{21cm,LAE}$ for the differential 21cm brightness temperature and LAE distribution as a function of distance $r$ from galaxies. In each panel, lines indicate $\xi_{21cm,LAE}$ for different IGM ionization states, ranging between \avchi$=0.9$ - $10^{-4}$; shaded regions show the uncertainties associated with idealized SKA1-Low and Subaru Hyper Suprime Cam 1000~h observations. As marked, panels show results for different $f_{esc}$ values ranging between $0.05$ and $0.95$. As shown, these combined observations can provide exquisite constraints on the IGM ionization state, allowing us to differentiate between \avchi $\sim$ 0.5, 0.25, 0.1 and 0.01 (see Sec. \ref{subsec_21cm_LAE_crosscorrelation} for details). \label{fig_21cm_LAE_crosscorrelation}}
\end{figure*}

These results are in general agreement with those obtained by other groups \citep{lidz2009,wiersma2013,park2014,vrbanec2015, sobacchi2016}, in terms of finding that whilst 21cm emission and galaxies are anti-correlated on small scales, they are uncorrelated on the largest scales. It is reassuring to find these results given the differences compared to previous works: we use a full 3D radiative transfer calculation that accounts for the spatially dependent clumping factor in each grid cell down to the resolution scale of the hydrodynamical simulation. On the other hand, \citet{lidz2009} and \citet{park2014} employ a semi-numerical scheme, which assumes a cell to attain $\chi_{HII}=1$ as soon as the ionization criterion are met; their results are therefore insensitive to low $\chi_{HI}$ values inside ionized regions. In contrast, while \citet{wiersma2013} solve the radiative transfer equation, they reduce the problem to 1D by assuming a spherically averaged density profile for each source for which reason their ionization fields are insensitive to 3D density inhomogeneities. The calculations carried out by \citet{vrbanec2015} are most comparable to ours in terms of the radiative transfer. However, they assume the ionizing photon rate to simply scale with the halo mass such that $Q=9.5\times10^{52} (M_h/10^{10}\msun)$. In contrast, we use the age and metallicity dependent output of ionizing photons from all star particles in bound galaxies in this work, resulting in $M_h<2\times10^{10}\msun$ showing ionizing emissivities ranging over 5 orders of magnitude ($Q\simeq 10^{49}-10^{54}$s$^{-1}$). Naturally, this means galaxies with the lowest ionizing photon outputs are unable to ionize any significant volume around themselves. Perhaps the most crucial difference is that \citet{vrbanec2015} only consider massive halos with halos with $M_h>10^{10}\msun$ in their 21cm-galaxy cross correlation calculations, while we include galaxies that an order of magnitude less massive. 

%***************************************************************************************************
\subsection{The 21cm-LAE cross-correlation}
\label{subsec_21cm_LAE_crosscorrelation}
%***************************************************************************************************
We investigate in Fig. \ref{fig_21cm_LAE_crosscorrelation} the cross correlation between the 21cm signal and the subset of galaxies visible as LAEs. Unlike the entire galaxy population used in the previous section, the visibility of galaxies as LAEs sensitively depends on $f_{esc}$ and $\langle \chi_{HI} \rangle$ - both the 21cm and LAE fields therefore evolve with these two parameters. We now show the 21cm-LAE cross-correlation function $\xi_{21cm, LAE}$ for all $f_{esc}$ and $\langle \chi_{HI} \rangle$ values in agreement with the observed Ly$\alpha$ LFs.

Firstly, we find that there is no-correlation ($\xi_{21cm, LAE}=0$) for \avchi$\lsim 0.01$, for all $f_{esc}$ ranging between $5\%$ to $50\%$, due to the lack of any 21cm emission. For higher \avchi~ values, $\xi_{21cm, LAE}$ shows a clear anti-correlation on small ($\lsim 20 h^{-1}$ cMpc) scales with an amplitude that decreases (i.e. anti-correlation weakens) from $-0.25$ to $-0.05$ as the IGM decreases from being $50\%$ to $10\%$ neutral - this trend is essentially driven by a decrease in the amplitude of the 21cm power spectrum as the IGM becomes progressively ionized. This anti-correlation is much more pronounced than that seen for the 21cm signal and the entire underlying galaxy population as shown in Sec. \ref{subsec_21cm_gal_crosscorrelation} above.
This is because only galaxies in predominantly clustered regions, hosting the most luminous galaxies with stellar masses $\gsim 10^{9.5}\Msun$ \citep[see e.g.][]{hutter2015}, lie in sufficiently large ionized regions to transmit enough Ly$\alpha$ luminosity to be visible as LAEs.
As a result of their large masses, and therefore ionizing photon output, LAEs comprise the galaxy subset that reside in the most ionized (and over-dense) regions of the simulation (see Fig. \ref{fig_dens_XHI}) where all the \HI is ionized. As expected, the 21cm and LAE distribution are uncorrelated at scales larger than $20 h^{-1}$ cMpc. 

Further, we find that although $\xi_{21cm, LAE}$ shows similar qualitative trends for all the three $f_{esc}$ values, its value becomes more negative, showing a stronger anti-correlation, as $f_{esc}$ increases from $5$ to $50\%$. This is because an increasing $f_{esc}$ value results in a higher emissivity and hence a lower \HI content, resulting in a lower 21cm emission near LAEs.  

We then calculate the joint SKA1-Low-Subaru Hypersuprime Cam (HSC) detectability of the 21cm-LAE correlation. The $1\sigma$ error bars displayed in Fig. \ref{fig_21cm_LAE_crosscorrelation} are computed for an idealised Subaru HSC and 1000~h SKA1-Low experiment. The errors include thermal noise for a 1000~h SKA experiment, and sample variance for both optical and radio measurements. We assume the same SKA1-Low configuration and system specifications as in the previous Section. However, since the field-of-view (FOV) of Subaru is smaller than for SKA, we assume the observation volume to be limited by Subaru. This reduces the number of independent samples within the volume, and the sample variance increases relative to an SKA-only experiment. The Subaru Suprime camera specifications are assumed, with a FOV of 34 arcmin x 27 arcmin. 
Again we match the bandwidth to each distance, $r$, in the two-point cross-correlation function, with a minimum resolution of 1.9$h^{-1}$~Mpc. 
We assume that the Subaru narrowband filter can resolve scales of this size, although, in practice, the NB921 filter has an intrinsic resolution of $\sim7h^{-1}$~Mpc at $z=6.6$. Convolution of the simulated cube with this narrowband filter (i.e. taking the narrowband resulting redshift uncertainties into account) would suppress small scale power, but this is not considered in this work. The corresponding decrease in the cross correlation amplitude will be approximately the ratio between the effective spectral depth for a given separation r and the Subaru spectral resolution. However, the increasing number of spectroscopically followed-up LAEs will allow us to exploit more precise redshift measurements of LAEs.
Under these assumptions, the synergistic SKA1-Low-HSC experiment would be able to yield constraints on the IGM ionization state. 
{\it Indeed, as shown in the Fig. \ref{fig_21cm_gal_crosscorrelation}, we find that the SKA1-Low-HSC experiment would be able to distinguish between \avchi~of $10\%$, $25\%$ and $50\%$, in addition to being able to differentiate a 10\% neutral IGM from one which was fully ionized, irrespective of the parameter space ($f_{esc}$, $f_\alpha/f_c$) explored.}

Including an evolution of the galaxy population and density contrast should not have a considerable impact on the 21cm-LAE cross correlation function on scales $<10 h^{-1}$Mpc considered here: Firstly, the propagation speed of the ionization fronts is about $1$~Mpc in $\Delta z \simeq0.4$ for the mean galaxy and even higher for LAEs; thus the photoionization rate in the vicinity of galaxies is largely determined by the present source. Secondly, LAEs are luminous galaxies residing in clustered regions (most overdense and ionized); due to their similar properties, the photoionization and thus ionization field in their vicinity should be comparable, and the amplitude of the anti-correlation should remain similar for same \avchi~values. The same rationale applies to observed galaxies. However, an evolving galaxy population introduces continuously partially ionized cells, and a growing density contrast keeps the gas density imprinted in the residual \HI; both lead to a rather constant amplitude of the oscillations in the 21cm-galaxy cross correlation functions.

Increasing the spatial resolution of our simulations would allow us to resolve Lyman Limit systems, which will damp the emitted Ly$\alpha$ luminosities from adjacent galaxies, in particular luminous galaxies \citep{kakiichi2016}.  Their impact on the 21cm-LAE cross correlation functions depends on their \HI volume and content: while an increasing \HI volume leads to an overall boost of the 21cm brightness temperature on small scales of the cross correlation, a rising \HI content can damp the Ly$\alpha$ emission of adjacent galaxies just as much that those cannot be seen as LAEs anymore and do not contribute to the 21cm-LAE cross correlations. An analysis which of those two effects dominates is subject of future work.

\begin{figure*}
 \center{\includegraphics[width=0.9\textwidth]{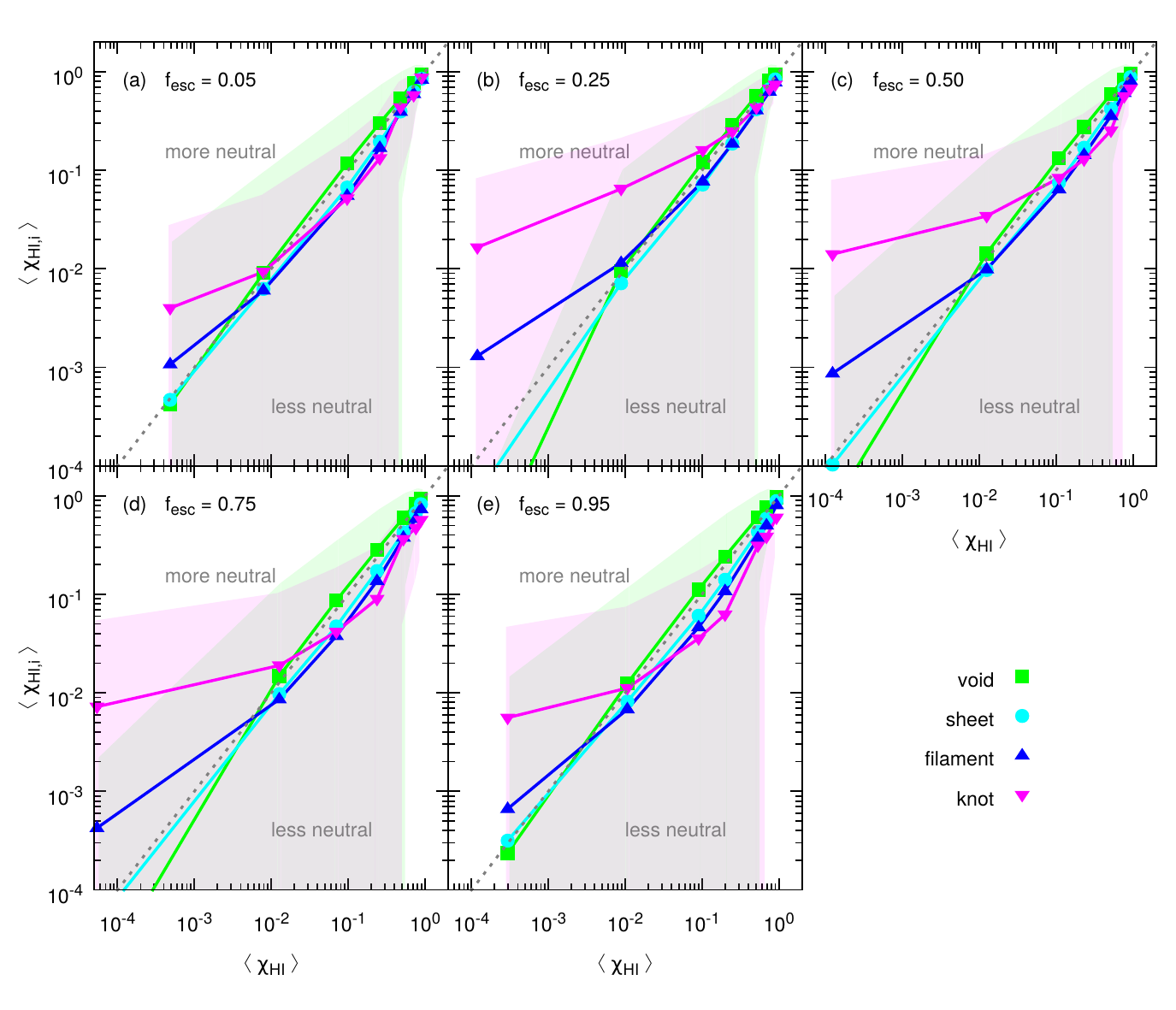}}
 \caption{Evolution of the mean neutral hydrogen fraction $\langle\chi_{HI,i}\rangle$ in the cosmic web components - voids (green squares), sheets (light blue circles), filaments (dark blue upside triangles) and knots (magenta downside triangles) - as a function of the overall mean neutral hydrogen fraction \avchi~for the $f_{esc}$ value marked in each panel. The gray dotted line indicates $\langle\chi_{HI,i}\rangle=$\avchi. The $1-\sigma$ standard deviations for knots and voids are indicated by correspondingly colored areas. \label{fig_cosmicweb_ionhist}}
\end{figure*}

% ***************************************************************************************************
\section{The topology of reionization}
\label{sec_reion_topology}
% ***************************************************************************************************
The reionization topology remains a much studied topic with theoretical approaches ranging from relating the mass to the volume averaged ionization fraction \citep{iliev2006a} to using Minkowski functionals \citep{gleser2006, lee2008, friedrich2011} to computing cross correlations of density and ionization redshift fields \citep{battaglia2013}. These have yielded results ranging from the popular ``inside-out" scenario where densest regions close to sources are ionized first with under-dense filaments being ionized last \citep{iliev2006a, iliev2007, trac2007, dayal2011a, battaglia2013, bauer2015} to the ``outside-in" topologies that predict the opposite \citep{miralda-escude2000}. We start by determining the reionization topologies in the different cosmic web components (knots, filaments, sheets and voids) in Sec. \ref{subsec_ionhist_cosmicweb}. We then build on the analytic approach proposed by \citet{wyithe2007} to examine the signatures of 21cm emission in over-densities (hosting galaxies) and voids in Sec. \ref{subsec_measurement}, in order to shed light on whether reionization proceeded ``inside-out", or vice-versa.

% ***************************************************************************************************
\subsection{Ionization history of the cosmic web}
\label{subsec_ionhist_cosmicweb}
% ***************************************************************************************************

We start by classifying the cosmic web into knots, filaments, sheets and voids, following a slightly modified approach to the tidal field tensor method proposed by \citet{hahn2007}. We first calculate the tidal field tensor 
\begin{eqnarray}
T_{ij}&=& \frac{\partial \Phi}{\partial x_i\ \partial x_j},
\end{eqnarray}
and compute the three eigenvalues $\lambda_i$, which quantify the curvature of the gravitational potential $\Phi$. \citet{hahn2007} propose classifying structures collapsing ($\lambda_i>0$) along three, two and one spatial dimensions as knots, filaments and sheets, respectively; structures showing no collapse in any dimension ($\lambda_i<0$) are classified as voids. However, \citet{forero-romero2009} have pointed out that such a scheme results in a very low volume filling fraction for voids. This is because (infinitesimally) small positive eigenvalues represent a scenario wherein the collapse will occur in the distant future; inspection at the present time would therefore not classify these regions as collapsing. To correct for this, we classify structures according to the {\it number} of eigenvalues ($N_{\lambda}$) above a threshold ($\gamma$) - cells with $N_{\lambda}=3$, $2$, $1$, $0$ are identified as knots, filaments, sheets and voids, respectively. In our calculations we use a threshold of $\gamma=0.3$, resulting in $60$\% of the volume being identified as voids. The cosmic web and its classification on the threshold ($\gamma$) from our simulated volume are shown in Appendix \ref{a1_sec_cosmicweb}.

In Fig. \ref{fig_cosmicweb_ionhist} we show the average \HI fractions for the four cosmic web components: voids ($\langle\chi_{HI,v}\rangle$), sheets ($\langle\chi_{HI,s}\rangle$), filaments ($\langle\chi_{HI,f}\rangle$), and knots ($\langle\chi_{HI,k}\rangle$) for all the $f_{esc}$ and reionization states values used in this work. The grey dotted line marks $\langle\chi_{HI,i}\rangle=\langle\chi_{HI}\rangle$, i.e. ionization values lying above and below this line imply structures that are less and more ionized than the average IGM ionization state, respectively.

For an IGM more neutral than $\langle \chi_{HI} \rangle \gsim 0.1$, we find the \HI fraction to be the lowest in knots, followed by filaments, sheets and voids in that order, i.e. $\langle\chi_{HI,k}\rangle<\langle\chi_{HI,f}\rangle<\langle\chi_{HI,s}\rangle<\langle\chi_{HI,v}\rangle$. This increase in the \HI fraction from over-dense to under-dense regions shows that reionization follows the ``inside-out scenario" where ionization fronts propagate from (galaxies in) the densest regions, reaching the most under-dense voids last. Further the continual output of ionizing photons ensures the densest regions remain ionized, at least in the initial reionization stages, given the long recombination timescales. The situation reverses (undergoes an ``inversion") in the end stages of reionization when $\langle \chi_{HI} \rangle \lsim 10^{-2}$, such that voids are now the most ionized, followed by sheets, filaments and knots, respectively. This arises as a result of the largest gas densities pushing up the average $\chi_{HI}$ values in the most over-dense knots; ionization fronts from multiple sources and lower gas-densities ensure a lower neutral fraction in under-dense voids. Naturally, such a behavior is expected to arise only at the end stages of reionization when ionized regions around galaxies essentially percolate throughout the IGM, leading to a more homogeneous photoionization rate.

As for $f_{esc}$, this parameter affects both the volume ionized by a source as well as the ionization fraction within it. As noted, a degeneracy exists between these two quantities such that, for a given $\langle \chi_{HI}\rangle$, galaxies build smaller ionized volumes containing larger ionized fractions over smaller timescales for an increasing $f_{esc}$. This naturally leads to a shift in the ionization fraction of the cosmic web components - as $f_{esc}$ increases from 0.25 to 0.95, the inversion occurs at successively lower $\langle \chi_{HI}\rangle$ values. As expected, given that the photoionization rate drops with the square of the distance from the source, the strongest effect of $f_{esc}$ is felt by over-dense knots. Finally, $f_{esc}=0.05$ represents a special case since the long timescales of about 1 Gyr required to reionize the IGM in this case result in recombinations becoming important, specially in the densest regions. Finally, we have carried out these calculations for various values of the Eigenvalue threshold ($\gamma$) ranging between 0. and 0.7 to ensure that our qualitative results are independent of the precise value used.

Our results, therefore, support a scenario where reionization proceeds from over-dense to under-dense regions, in agreement with most other works \citep[e.g.][]{trac2007,iliev2012,battaglia2013,bauer2015}. However, our results are in tension with \citet{finlator2009} who find filaments to be ionized last. We discuss some possible reasons for this disagreement: first, whilst our RT resolution is lower than theirs, we properly account for the local clumping factor for each cell, down to the resolution of the SPH simulation. Secondly, we find the emissivity bias for galaxies with the youngest stellar populations to be lower than that found by \citet{finlator2009}, where the ionizing emissivity scales with halo mass as $M_h^{1.3}$. 

% ***************************************************************************************************
\subsection{21cm emission from regions with and without galaxies}
\label{subsec_measurement}
% ***************************************************************************************************

\begin{figure*}
 \center{\includegraphics[width=0.9\textwidth]{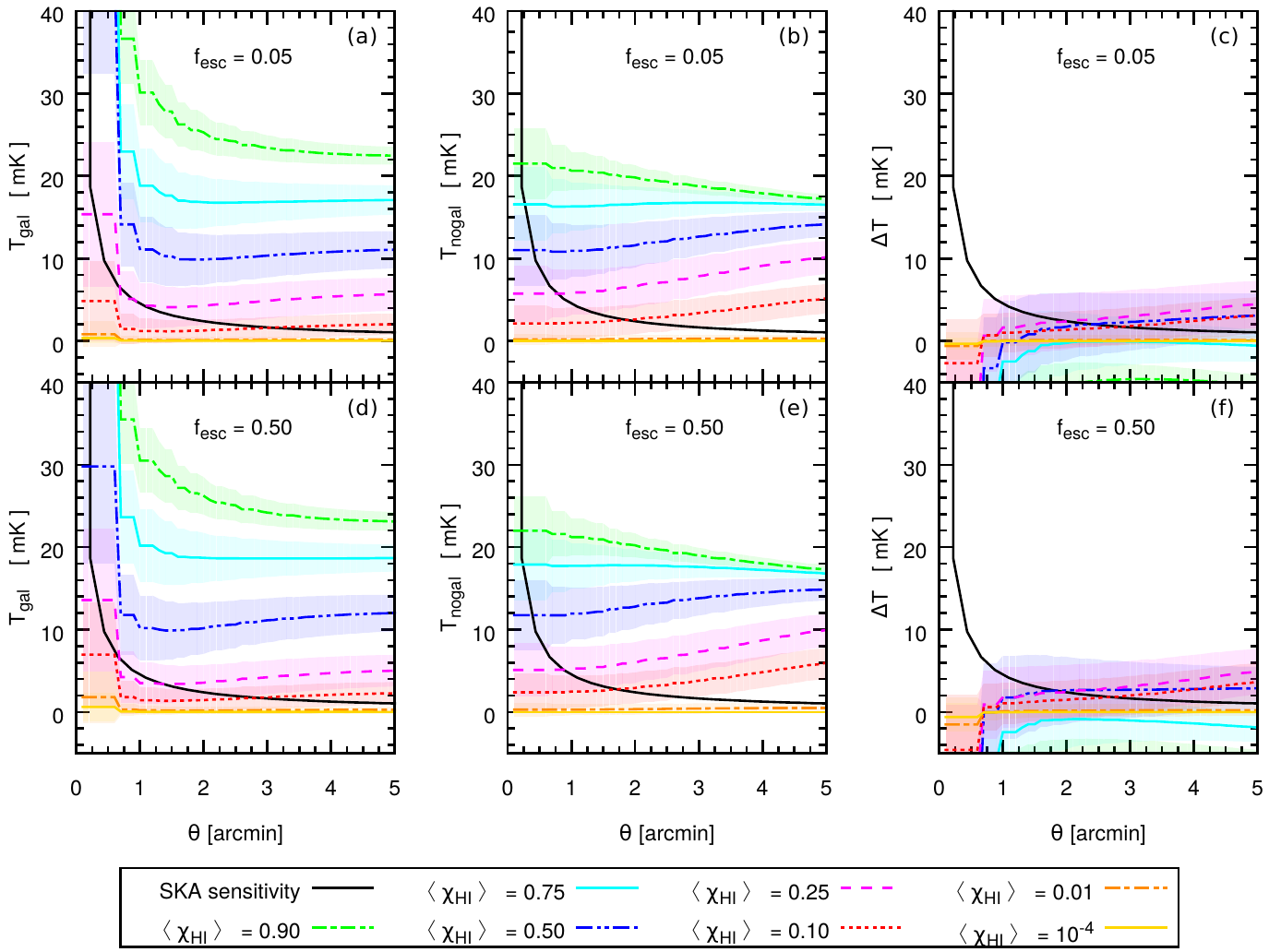}}
 \caption{Differential 21cm brightness temperature in regions containing galaxies (left column), in regions not containing galaxies (central column), and their difference $\Delta T =  T_{\mathrm{nogal}}-T_{\mathrm{gal}}$ (right column) as a function of the smoothing scale $\theta$. The upper and lower panels show the results for $f_{esc}=0.05$ and $0.5$, respectively. In each panel, we show the differential brightness temperature at different stages of reionization (\avchi$=0.9$ - $10^{-4}$); the solid black line shows the SKA imaging sensitivity limits for a 1000~h observation. \label{fig_21cm_diff_gal_nogal_summary}}
\end{figure*}

As shown in Section \ref{subsec_gal_tracer}, galaxies and in particular LAEs, reside in over-dense and highly ionized regions, thereby exhibiting a lower 21cm brightness temperature as compared to similarly over-dense regions devoid of galaxies. As the Universe approaches complete ionization, the ionizing emissivity (and hence \avchi) becomes more homogeneous and causes the \HI distribution to follow the underlying spatial density distribution; the 21cm brightness temperature in under-dense regions then drops below that in over-dense regions. We now investigate the detectability of the reionization topology by combining 21cm emission information with galaxy surveys \citep[see also][]{wyithe2007}.

We start by dividing the 21cm differential brightness temperature calculated in equations \ref{eq_21cm_brightness_temperature} and \ref{eq_T0} into cells that contain galaxies as
\begin{eqnarray}
 T_{\mathrm{gal}}&=& T_0\ \langle \chi_{\mathrm{HI}} \rangle\ \langle \left( 1+\delta_{\mathrm{HI}}(\vec{x}) \right) \left( 1+\delta(\vec{x}) \right) \rangle_{\vec{x}\in\mathrm{V_{gal}}},
\end{eqnarray}
while $\delta T_b$ is defined as the temperature in cells not containing galaxies:
\begin{eqnarray}
 T_{\mathrm{nogal}}&=& T_0\ \langle \chi_{\mathrm{HI}} \rangle\ \langle \left( 1+\delta_{\mathrm{HI}}(\vec{x}) \right) \left( 1+\delta(\vec{x}) \right) \rangle_{\vec{x}\in\mathrm{V_{nogal}}}.
\end{eqnarray}
To imitate the observational angular resolution of any imaging (say $\theta$) we convolve these temperatures with a top-hat filter of width $\Delta s=1.8 h^{-1}\mathrm{cMpc}\ (\theta/\mathrm{arcmin})^{-1}$, corresponding to the comoving distance of two points separated by an angular distance $\theta$ on the sky at $z\simeq6.6$. 
From the convolved temperature fields we derive the variances for $T_{\mathrm{gal}}$ and $T_{\mathrm{nogal}}$, whereas the corresponding volumes depend on $\theta$. In an experiment, where $N$ independent fields of an angular size $\theta$ and corresponding volume $\pi(\Delta s)^3/6$ are measured, the variance reduces by $N^{-1/2}$. For $N=10$ the corresponding average values along with the $1\sigma$ errors are shown in Fig. \ref{fig_21cm_diff_gal_nogal_summary} for two representative cases of $f_{esc} =0.05$ and $0.5$. 
In order to compare to observations, we calculate the SKA1 imaging sensitivity using the same SKA1-Low array configuration (V4A configuration) as was used in Section \ref{sec_link_21cm_gal}, comprising a densely-packed core and outer stations configured in a spiral-like configuration. The computation considers two polarizations, a $1000$~h observation, and a $1~$MHz bandwidth. 
We propose to conduct a $1000$~h SKA1 observation of such angular size that $10$ non-overlapping fields with angular size $\theta$ containing (not containing) galaxies can be extracted.

\begin{figure*}
 \center{\includegraphics[width=0.9\textwidth]{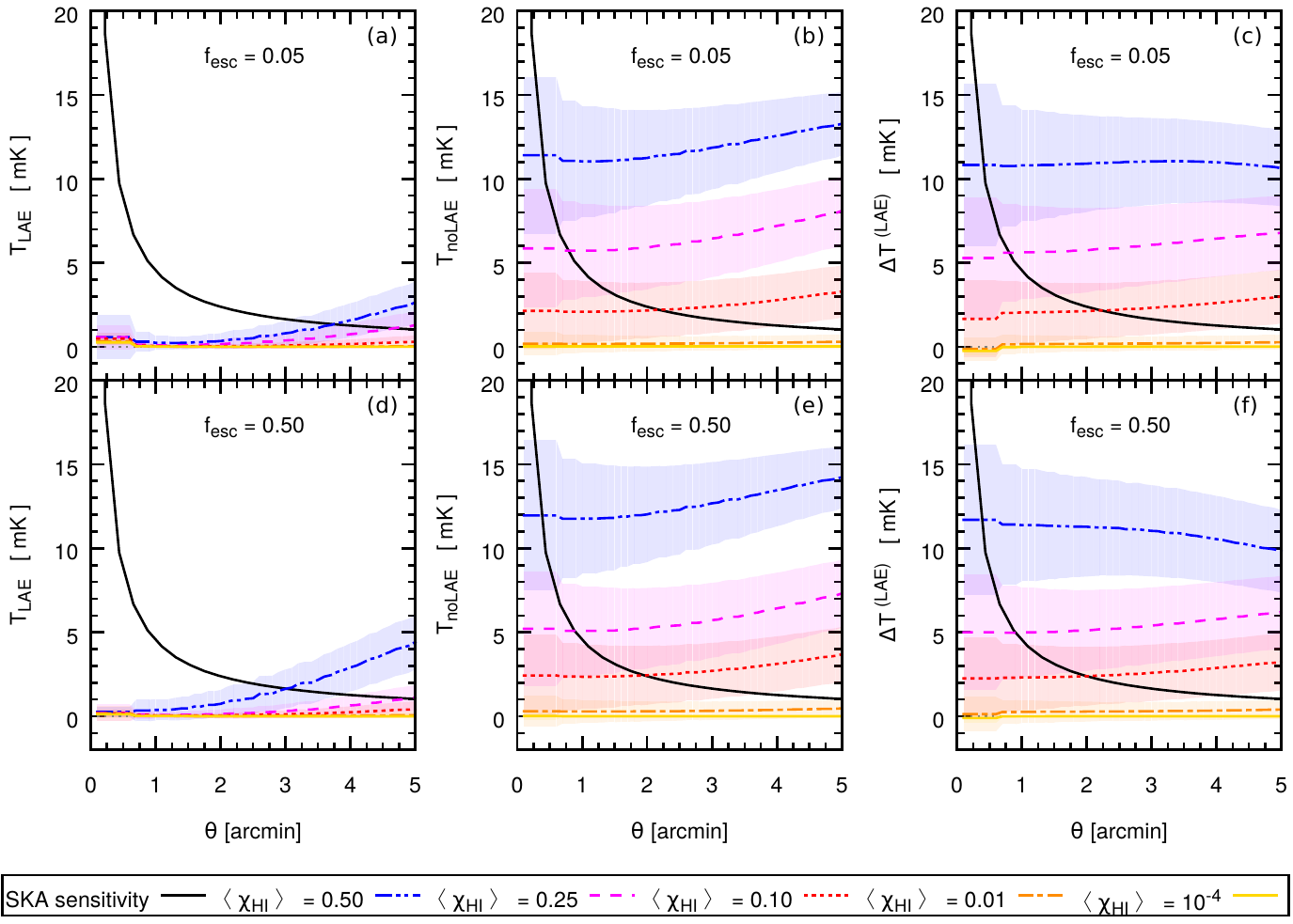}}
 \caption{Differential 21cm brightness temperature in regions containing LAEs (left column), in regions not containing LAEs (central column), and their difference $\Delta T =  T_{\mathrm{noLAE}}-T_{\mathrm{LAE}}$ (right column) as a function of the smoothing scale $\theta$. The upper and lower panels show the results for $f_{esc}=0.05$ and $0.5$, respectively. In each panel, we show the differential brightness temperature at different stages of reionization (\avchi$=0.9$ - $10^{-4}$); the solid black line shows the SKA imaging sensitivity limits for a 1000~h observation. SKA-LOW1 should be able to detect the difference between an IGM ionized at the 10\% or 50\% level focusing on fields without LAEs (panels b,e). The higher 21cm brightness temperature in low-density regions devoid of galaxies also provides support for the ``inside-out" topology of reionization. \label{fig_21cm_diff_LAE_noLAE_summary}}
\end{figure*}

Then, a comparison of the 21cm brightness temperature ($\Delta T$) in regions with and those without galaxies provides an estimate of the reionization topology, allowing constraints on whether reionization had progressed faster in over-dense regions (hosting galaxies) or under-dense regions (devoid of galaxies): 
\begin{eqnarray}
\Delta T =  T_{\mathrm{nogal}}-T_{\mathrm{gal}}.
 \label{eq_Tgal_Tgalint}
\end{eqnarray}

In the initial reionization stages ($\langle \chi_{HI} \rangle \simeq 0.9$), a number of low-mass ($M_{\star}\lesssim10^{9.5}\Msun$) field galaxies are embedded in only partially ionized regions at scales $\lsim 1$~arcmin ($\sim2.5$~cMpc) - this combined with the high gas densities around galaxies results in brightness temperatures as high as $T_{\mathrm{gal}}\gsim 40$~mK as seen from panel (a) of Fig. \ref{fig_21cm_diff_gal_nogal_summary}, where $f_{esc}=0.05$. The progress of reionization leads to a drop in the \HI content even around low-mass field galaxies, resulting in a drop in $T_{\mathrm{gal}} \sim 16$~mK for $\langle \chi_{HI} \rangle \simeq 0.25$ at these scales. Finally, $T_{\mathrm{gal}}$ drops to about 0~mK once the IGM is more ionized than 99\%, or $\langle \chi_{HI} \rangle \lsim 0.01$. At increasing angular scales, $T_{\mathrm{gal}}$ drops due to a decrease in the gas density - this is mostly driven by the lower gas density in voids that cover roughly 60\% of the simulation volume. In this case too, $T_{\mathrm{gal}}$ scales with the average IGM ionization state, dropping from $\sim 22$~mK to $\sim 2$~mK as $\langle \chi_{HI} \rangle$ decreases from 0.9 to 0.1 for $\theta \simeq 4$~arcmin ($\simeq 10.2$~cMpc); as expected $T_{\mathrm{gal}} \sim 0$~mK once $\langle \chi_{HI} \rangle$ drops below 0.01.

$T_{\mathrm{nogal}}$ shows lower temperatures as a result of the lower gas-densities in regions devoid of galaxies.
This trend may seem counterintuitive, however we remind the reader that $T_{gal}$ is strongly driven by the high gas densities galaxies reside in, in particular those galaxies in partially ionized cells.
$T_{\mathrm{nogal}}$ still scales with the IGM ionization state, decreasing from $\sim 22$~mK to $\sim 2$~mK as $\langle \chi_{HI} \rangle$ drops from $0.9$ to 0.1 at 1 arcmin scales. Given that $T_{\mathrm{nogal}}$ probes lower gas-density contrasts in regions devoid of galaxies, its scale variation is less than that seen for $T_{\mathrm{gal}}$: $T_{\mathrm{nogal}}$ varies by $\sim 6$~mK from scales ranging between $0.1$ to 5~arcminutes, compared to $T_{\mathrm{gal}}$ that can vary by as much as 20~mK (for $\langle \chi_{HI} \rangle\gsim 0.5$) on the same scales. Finally, we note that $T_{\mathrm{nogal}} \sim 0$~mK on all scales once $\langle \chi_{HI} \rangle$ drops below 0.01.

As expected from the above discussion, the temperature difference between regions without and with galaxies ($\Delta T$) is negative at scales less than about 0.8~arcmin, where $T_{\mathrm{gal}}$ is enhanced as a result of (low-mass field) galaxies being embedded in high-density, partially ionized regions. $\Delta T$ naturally flips in sign, becoming slightly positive ($\sim 2-4$~mK) at larger scales for $\langle \chi_{HI} \rangle \lsim 0.5$; as expected $\Delta T \sim 0$~mK for an IGM more ionized than 99\%. These trends remain the same even if the ionizing photon escape fraction increases by an order of magnitude to $f_{esc}=0.5$, as shown in the lower 3 panels of Fig. \ref{fig_21cm_diff_gal_nogal_summary}.

In terms of observability, conducting a 1000~hr survey including 10 fields around regions with galaxies, SKA1 should be able to distinguish between IGM ionization states of $\langle \chi_{HI} \rangle \simeq 0.25$, 0.5, 0.75 and 0.9 at scales greater than 1~arcminute, irrespective of the $f_{esc}$ value ranging between 5\% or 50\%; the brightness temperatures for an IGM more ionized than 90\% are too close to the SKA1 detection limits to be unambiguously identified. Centering the beam on regions devoid of galaxies, SKA1 should be able to distinguish between an IGM more or less ionized than 25\% for beams larger than 2~arcminutes. However within error bars, the difference in temperature ($\Delta T$) between regions with and without galaxies is too close to the SKA detection limits to be able to constrain the nature of reionization (inside-out or outside-in).

% ***************************************************************************************************
\subsection{21cm emission from regions with and without LAEs}
\label{subsec_measurement2}
% ***************************************************************************************************
In a next step we calculate the 21cm differential brightness temperature in cells that contain LAEs as
\begin{eqnarray}
 T_{\mathrm{LAE}}= T_0\ \langle \chi_{\mathrm{HI}} \rangle\ \langle \left( 1+\delta_{\mathrm{HI}}(\vec{x}) \right) \left( 1+\delta(\vec{x}) \right) \rangle_{\vec{x}\in\mathrm{V_{LAE}}},
\end{eqnarray}
while $\delta T_b$ for cells not containing LAEs is calculated as
\begin{eqnarray}
 T_{\mathrm{noLAE}}= T_0\ \langle \chi_{\mathrm{HI}} \rangle\ \langle \left( 1+\delta_{\mathrm{HI}}(\vec{x}) \right) \left( 1+\delta(\vec{x}) \right) \rangle_{\vec{x}\in\mathrm{V_{noLAE}}}.
\end{eqnarray}
Finally, the difference in the 21cm brightness temperature in regions with/without LAEs can be expressed as
\begin{eqnarray}
\Delta T^{(LAE)} = T_{\mathrm{noLAE}} - T_{\mathrm{LAE}},
\end{eqnarray}
results for which are shown in Fig. \ref{fig_21cm_diff_LAE_noLAE_summary}. We note that $T_{\mathrm{noLAE}}$ may contain non-Ly$\alpha$ emitting galaxies.

We start by noting that we only match the observed LAE LFs for $\langle \chi_{HI} \rangle \simeq 0.5$ which marks the upper limit for both $T_{\mathrm{LAE}}$ and $T_{\mathrm{noLAE}}$. Given that LAEs represent the subset of galaxies located in the most over-dense and ionized regions (see Sec. \ref{subsec_gal_tracer}), we find $T_{\mathrm{LAE}} \sim 0$~mK at all scales, for both $f_{esc}=0.05$ and 0.5, for $\langle \chi_{HI} \rangle \lsim 0.5$ as shown in Fig. \ref{fig_21cm_diff_LAE_noLAE_summary}. For $\langle \chi_{HI} \rangle \simeq 0.5$, on the other hand, $T_{\mathrm{LAE}}$ shows a slight increase in temperature from 0 to $\sim 5$~mK with increasing scale (from $0.1$ to $5$~arcminutes), since we effectively sample the brightness temperatures of voids at such large scales. We remind the reader that an average IGM ionization state of $\langle \chi_{HI} \rangle \simeq 0.5$ is obtained due to a higher ionized fraction inside smaller ionized volumes, as $f_{esc}$ increases from 0.05 to 0.5. Naturally, the smaller total ionized volume results in a larger neutral fraction, boosting $T_{\mathrm{noLAE}}$ to about 12~mK at the largest scales for $f_{esc}=0.5$. 

Given LAEs occupy the largest halos in the most ionized regions, $T_{\mathrm{noLAE}}$ is generically higher than $T_{\mathrm{nogal}}$ and shows a steady decrease as the IGM becomes increasingly ionized: $T_{\mathrm{noLAE}}$ falls from $\sim 12$ to 2.5~mK as $\langle \chi_{HI} \rangle$ decreases from 0.5 to 0.1 at $\sim 0.1$~arcminute scales; again, $T_{\mathrm{noLAE}} \sim 0$~mK, if the IGM is more ionized than 99\%. Finally, given that voids are more neutral in the initial reionization stages (see Sec. \ref{subsec_ionhist_cosmicweb}), $T_{\mathrm{noLAE}}$ increases by about 2~mK as $\theta$ increases from 0.1 to 5~arcminutes. Finally, we note that $T_{\mathrm{noLAE}}>T_{\mathrm{LAE}}$ results in a positive value of $\Delta T^{(LAE)}$ at all scales for $\langle \chi_{HI} \rangle > 0.01$; $\Delta T^{(LAE)} \simeq 0$~mK for a more ionized IGM. 

In terms of SKA observations, we find that SKA1 should be able to detect $T_{\mathrm{LAE}} \sim 2-4$mK values at scales greater than 3 arcminutes. With its larger values, $T_{\mathrm{noLAE}}$ provides a much cleaner probe: SKA1 should be able to distinguish between $\langle \chi_{HI} \rangle \simeq 0.1$ and 0.5 at scales larger than 2~arcminutes, irrespective of $f_{esc}$: this naturally implies {\it $\Delta T^{(LAE)}$ can also be used to differentiate between an IGM that is 10\% neutral to one that is 50\% ionized at these scales}. The fact that $\Delta T^{(LAE)}>0$~mK and $T_{\mathrm{LAE}}\simeq0$~mK therefore support the inside-out scenario, where ionized regions percolate from over- to under-dense regions in the IGM, characterized by a lower differential 21cm brightness temperature in regions around LAEs compared to regions not containing galaxies. This provides a promising experiment for combining future LAE Subaru and 21cm SKA observations \citep[see also][]{wyithe2007}.

Including an evolution of the galaxy population and the gas density increases not only the 21cm brightness temperature in neutral regions but also leads to a decrease in the density contrast towards higher redshifts (rising \avchi~values). While the first effect will be dominant in underdense regions, the latter may balance the first in overdense regions, i.e. $T_{\mathrm{gal}}$ remains similar for small $\theta$ but can be larger at higher $\theta$. However, given the reduced optical depth as indicated by Planck, reionization progressed later and faster than anticipated, by what the mentioned effects would become secondary.

An increase of the spatial resolution of our simulations could reveal more details about the environment of galaxies, in particular the signatures of overdense Lyman Limit systems could be studied. Assuming LLS are located in filaments, we would find $T_{\mathrm{gal}}$ - in particular close to galaxies - to rise as the \HI content in LLS increases. LLS will affect $T_{\mathrm{gal}}$ (and $T_{\mathrm{nogal}}$ on small scales), however, it remains an open question how much $T_{LAE}$ will be affected given the balance between a low enough \HI content for a sufficient Ly$\alpha$ transmission and a high enough \HI volume for a significant difference in the 21cm signal.

% ***************************************************************************************************
\section{Conclusions and discussion}
\label{sec_conclusions}
% ***************************************************************************************************
We post-process a GADGET-2 simulation snapshot at $z \simeq 6.6$ with a dust model and a RT code (pCRASH), which provide our framework for high-$z$ galaxies, and specially the subset visible as LAEs. We perform 5 RT simulations with pCRASH, each adopting a different $f_{esc}$ value (for all galaxies) between $0.05$ and $0.95$. Starting from a neutral IGM (\avchi$=1$), we run pCRASH until the Universe is completely ionized in each case. From the resulting ionization fields we derive the associated 21cm brightness temperature maps, and compute the 21cm-galaxy and 21cm-LAE cross correlations, the results of which are now summarized:

\begin{itemize}
\item Whilst galaxies are located in the most over-dense regions ($1+\delta \sim 1.5 - 15$), the subset visible as LAEs preferentially occupy the densest ($1+\delta \sim 2-15$) and most ionized regions ($\chi_{HI} \lsim 10^{-2}$). This naturally results in 21cm brightness temperatures an order of magnitude lower ($\sim 1$mK) in regions hosting LAEs as compared to similarly over-dense regions ($1+\delta\sim 3$) devoid of them.  

\item The 21cm-LAE anti-correlation (that increases with the increasing $\langle \chi_{HI} \rangle$) at small ($\lsim 10$cMpc) scales will provide an exquisite probe of the {\it average ionization state} at high-$z$: within errors, a 1000 hour joint SKA-Low1-Subaru Hypersuprime Cam (HSC) experiment will be able to distinguish between an IGM that was fully ionized to one that was $10\%$, $25\%$ or $50\%$ neutral, irrespective of the parameter space ($f_{esc}$, $f_\alpha/f_c$) explored. 

\item Even conducting a 1000~h survey of 10 fields around regions with galaxies, SKA1 should be able to distinguish between IGM ionization states of $\langle \chi_{HI} \rangle_{gal} \simeq 0.25$, 0.5, 0.75 and 0.9 at scales greater than 1~arcminute. However, given their larger masses, the 21cm temperature around LAEs effectively tends to 0 at almost all scales. 

\item In terms of the {\it reionization topology}, for an IGM more neutral than $\langle \chi_{HI} \rangle \gsim 0.1$, we find the \HI fraction to be the lowest in knots, followed by filaments, sheets and voids in that order supporting the ``inside-out scenario". If fields devoid of LAEs can be identified, a SKA1 1000~h survey of 10 fields around regions with and without LAEs can be used as a probe of the reionization topology. A positive differential 21cm brightness temperature in voids that tends to 0 in regions hosting LAE at scales larger than 2~arcminutes will provide strong support for the inside-out reionization scenario.

\end{itemize}

We end by summarizing the major caveats in this work. As a natural consequence of simulating cosmological volumes, we do not resolve Lyman Limit systems (LLS). Including LLS can decrease the Ly$\alpha$ transmission along those LOS that traverse such systems, affecting the visibility of bright galaxies as LAEs \citep{kakiichi2016}. Secondly, the prevalence of LLS in large numbers at early cosmic epochs could, in principle, lead to knots/filaments being ionized last resulting in an inside-out-middle reionization (c.f. \citet{finlator2009}).

In our error estimate of the 21cm-LAE cross correlation in a combined Subaru HSC and 1000~h SKA experiment, we have assumed that the location of LAEs in the IGM can be measured with a maximum uncertainty of $1.9h^{-1}$~Mpc, whereas the redshift uncertainty from Subaru HSC filters will be about $7h^{-1}$Mpc. This may lead to increased uncertainties and a weaker anti-correlation signal in the 21cm-LAE cross correlations at scales $<7h^{-1}$Mpc. Furthermore, narrowband selected LAEs resemble more a 2D distribution, and resulting projection effect may weaken the cross correlation signal. However, we have chosen to (a) employ the full three-dimensional information to investigate trends of changing galactic properties, and (b) provide a rough estimate on the detectability with focus on SKA.

We also note that, due to poor constraints on its mass dependence, we assume the same escape fraction of ionizing photons for each galaxy. An evolution in $f_{esc}$ with mass and/or redshift could have implications for the progress and sources of reionization that might be reflected in the 21cm-galaxy/LAE correlations. Forthcoming observations with the James Webb Space Telescope will be invaluable on shedding light on this parameter to identify the sources of reionization. 

% ***************************************************************************
% ***************************************************************************
\acknowledgments

The authors would like to thank Marco Castellano, Dijana Vrbanec, Benedetta Ciardi, Darren Croton, Manodeep Sinha for useful discussions and comments. The authors acknowledge Peter Creasey for permission to use his python library as a basis for the computation of the power spectra, and Benedetta Ciardi for a collaboration in developing pCRASH. 
AH is supported under the Australian Research Council's Discovery Project funding scheme (project number DP150102987).
PD acknowledges support from the European Commission's CO-FUND Rosalind Franklin program. CMT is supported under Australian Research Council's Discovery Early Career Researcher funding scheme (project number DE140100316) and the Centre for All-sky Astrophysics (an Australian Research Council Centre of Excellence funded by grant CE110001020).
Finally, this research was supported by the Munich Institute for Astro- and Particle Physics (MIAPP) of the DFG cluster of excellence ``Origin and Structure of the Universe".

% \vspace{5mm}
% \facilities{}
% \software{}

\appendix

% ***************************************************************************************************
\section{Cosmic web}
\label{a1_sec_cosmicweb}
% ***************************************************************************************************

\begin{figure*}
 \center{\includegraphics[width=0.9\textwidth]{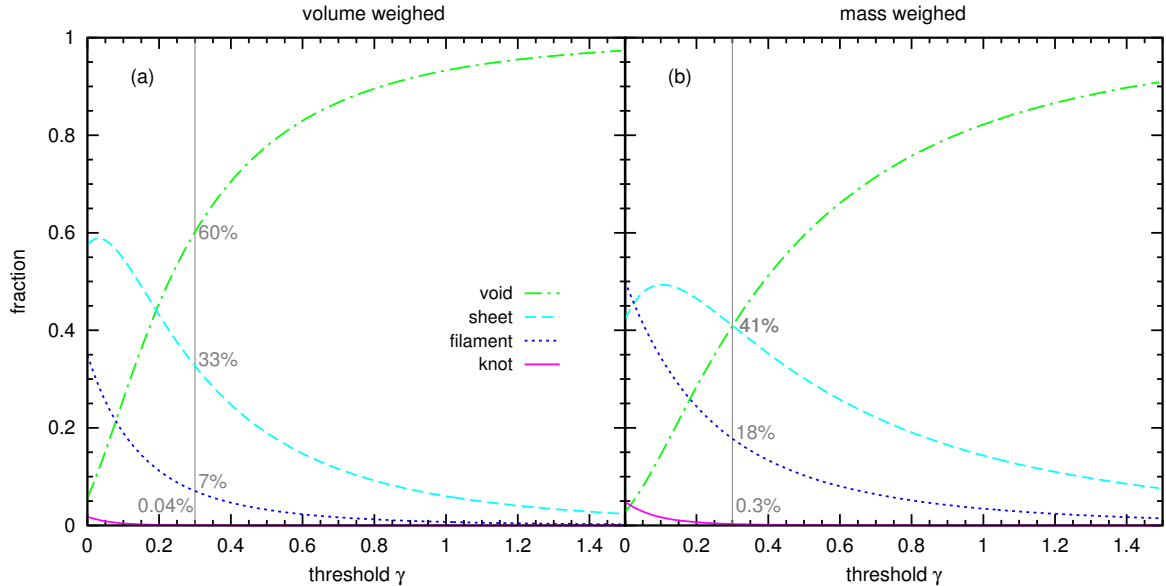}}
 \caption{Volume (left panel) and mass (right panel) weighed fractions of voids, sheets, filaments and knots versus the threshold $\gamma$ \citep{nuza2014}. The vertical line at $\gamma=0.3$ represents the threshold for which $60$\% of the volume is identified as voids and corresponds to the limit used in this work.  \label{fig_cosmicweb_threshold}}
\end{figure*}

In Fig. \ref{fig_cosmicweb_threshold} we show how the threshold $\gamma$ determines the classification of the cosmic web. An increase in the $\gamma$ (that can be thought of as an increase in the collapse time) results in an increase in both the mass and volume weighted fraction of voids, accompanied by a decrease in the values for knots, filaments and sheets. Given the mass concentration in knots, filaments and sheets - and the lack thereof in voids - we find that the mass weighted fraction of collapsed structures is always higher than the volume weighted fraction. 

\begin{figure}
 \center{\includegraphics[width=0.57\textwidth]{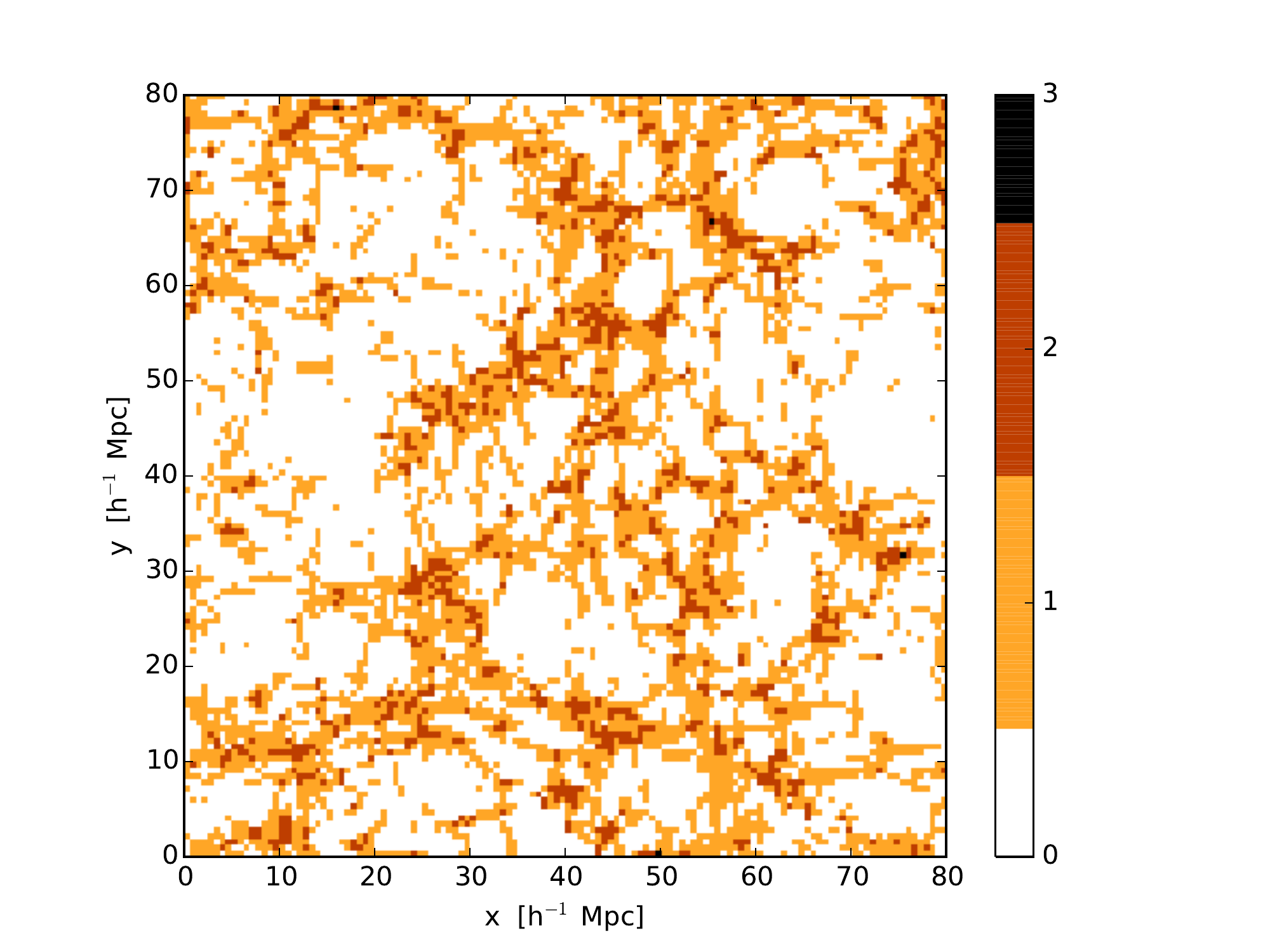}}
 \caption{A slice through the $80h^{-1}$cMpc simulation box showing the large-scale cosmic web. The different colors indicate the different cosmic web components: voids (white), sheets (orange), filaments (brown) and knots (black).\label{fig_cosmicweb_field}}
\end{figure}

In this paper we use a threshold that corresponds to a volume weighed fraction of 60\% for voids; this threshold is found at $\gamma=0.3$. The corresponding classification of the cosmic web is shown for a slice through the middle of the simulation in Fig. \ref{fig_cosmicweb_field}. 

% ***************************************************************************************************
\section{Results for a double peak Ly$\alpha$ line profile}
\label{a2_sec_lya_line_profile}
% ***************************************************************************************************

We compute the Ly$\alpha$ transmission $T_{\alpha}$ for a double peak line profile using a Gausian-minus-a-Gaussian (GmG) line shape with a width that depends on the halo mass as described \citet{jensen2013}. Their employed fitting function (equations 6-8 in \citet{jensen2013}) was obtained by fitting the emergent Ly$\alpha$ line profiles of a high-resolution galaxy sample spanning over a range of stellar masses. Assuming GmG line shapes, we find the Ly$\alpha$ transmissions $T_{\alpha}$ of our simulated galaxies boosted by $\sim0.1$ in comparison with $T_{\alpha}$ of Gaussian line shapes: the attenuation of Ly$\alpha$ radiation aside the resonance is the strongest, by what a larger fraction of the emergent Ly$\alpha$ radiation is attenuated for a Gaussian line profile. The boosted Ly$\alpha$ transmission can be compensated by a lower ratio between the escape fractions of Ly$\alpha$ and UV continuum photons ($f_{\alpha}/f_c$). We find that the population of galaxies visible as LAEs hardly changes, and thus the 21cm-LAE cross correlations do not change. In fact they look identical to the results in Fig. \ref{fig_21cm_LAE_crosscorrelation}.

\bibliographystyle{aasjournal}
\bibliography{crossps_apj_rev1}

\end{document}